\documentclass[3p]{elsarticle}

\usepackage{fancyhdr}
\usepackage{booktabs}
\usepackage{ctable}
\usepackage{amsmath}
\usepackage{mathtools}
\usepackage{commath}
\usepackage{graphicx}
\usepackage{nomencl}
\usepackage{commath}
\usepackage{natbib}
\usepackage{multirow}
\usepackage{longtable}
\usepackage{textcomp}
\usepackage{array}
\usepackage{graphics}
\usepackage{tabularx}
\usepackage{color}
\usepackage{geometry}
\usepackage{natbib}
\usepackage{mciteplus}
\usepackage{footnote}
\usepackage{amssymb}
\usepackage{pdflscape}
\usepackage{afterpage}
\usepackage{multirow}
\usepackage[title]{appendix}
\usepackage{subfigure}
\usepackage{caption}
\usepackage{amssymb}
\usepackage{natbib}
\usepackage{mleftright} 
\usepackage{latexsym}
\usepackage{color,soul}

\usepackage{etoolbox}
\renewcommand\nomgroup[1]{%
	\item[\bfseries
	\ifstrequal{#1}{A}{}{%
		\ifstrequal{#1}{G}{Greek Letters}{%
			\ifstrequal{#1}{S}{Subscripts}{}}}%
	]}

\usepackage[version=3]{mhchem} 

\usepackage{lineno,hyperref}

\journal{Advanced Materials Interfaces }

\biboptions{numbers,sort&compress}

\begin{document}
\begin{frontmatter}

\title{Ultrasensitive Surface\textendash Enhanced Raman Spectroscopy Detection by Porous Silver Supraparticles from Self\textendash lubricating Drop Evaporation}

\author{Tulsi Satyavir Dabodiya\textsuperscript{1,2}}
\author{Somasekhara Goud Sontti\textsuperscript{1}}
\author{Zixiang Wei\textsuperscript{1}}
\author{Qiuyun Lu\textsuperscript{1}}
\author{Romain Billet\textsuperscript{1}}
\author{Arumugam Vadivel Murugan\textsuperscript{2}}
\author{Xuehua Zhang\textsuperscript{*1,3}\corref{cor1}}
\ead{xuehua.zhang@ualberta.ca}
\address{1-Department of Chemical and Materials Engineering, University of Alberta, Alberta T6G 1H9, Canada}
\address{2-Centre for Nanoscience and Technology, Madanjeet School of Green Energy Technologies, Pondicherry University (A Central University), Kalapet, Puducherry 605014, India}
\address{3-Physics of Fluids Group, Max Planck Center Twente for Complex Fluid Dynamics, JM Burgers Center for Fluid Dynamics, Mesa+, Department of Science and Technology, University of Twente, Enschede 7522 NB, The Netherlands}

\newpage
\begin{abstract}	

		\noindent This work demonstrates an original and ultrasensitive approach for Surface\textendash Enhanced Raman Spectroscopy (SERS) detection based on evaporation of self\textendash lubricating drops containing silver supraparticles. The developed method detected an extremely low concentration of analyte that was enriched and concentrated on sensitive SERS sites of the compact supraparticles formed from drop evaporation. A low limit of detection (LOD) of $10^{-16}$ M was achieved for a  model hydrophobic compound rhodamine 6G (R6G). The quantitative analysis of R6G concentration was obtained from $10^{-5}$ M to $10^{-11}$ M.  In addition, for a model micro\textendash pollutant in water triclosan, the detection limit of $10^{-6}$ M  was achieved by using microliter sample solutions. The intensity of SERS detection in our approach was robust to the dispersity of the nanoparticles in the drop but became stronger after a longer drying time. The ultrasensitive detection mechanism is the sequential process of concentration, extraction, and absorption of the analyte during evaporation of self\textendash lubrication drop and hot spot generation for intensification of SERS signals. This novel approach for sample preparation in ultrasensitive SERS detection can be applied to the detection of chemical and biological signatures in areas such as environment monitoring, food safety, and biomedical diagnostics.
	
\end{abstract}


\end{frontmatter}

\footnote[1]{Equally contributing authors: Somasekhara Goud sontti and Zixiang Wei}

\newpage
\section{Introduction}
Surface\textendash enhanced Raman Spectroscopy (SERS) has attracted considerable attention because it can enhance weak Raman signals by several orders of magnitude and allow the detection of molecule fingerprints\cite{campion1998surface,jarvis2008characterisation} At present, SERS is a highly adaptive and versatile chemical analysis technique owing to its rich spectral information, capability for single molecular level detection, and non-destructive analysis nature.\cite{qian2008single,adu2012probing} SERS sensitivity enables it to trace chemical species down to a single molecular level. The most promising features of SERS have attracted increasing attention to its application in various fields such as environmental monitoring, food safety, biomedical diagnostics, gas sensors, and detection of explosives in defence systems.\cite{perumal2021towards,bantz2011recent,fan2020review,huang2020detection, sylvia2000surface, bharati2021flexible}


SERS detection comprises an electromagnetic and a chemical effect originating from the resonance Raman enhancement during specific metal\textendash molecule interactions. Three mechanisms contribute mainly to the enhancement of SERS signals under chemical conditions: charge transfer, molecular resonance, and non\textendash resonant interactions.\cite{morton2009understanding} The signal depends on the enhancement of electromagnetic fields in the hot spots of metal plasmonic nanoparticles via the plasmonic effect.\cite{baia2006surface,shiohara2020recent} The strong coupling of localized surface plasmon resonance (LSPR) on these hot spots provides a high enhancement factor to electromagnetic mechanisms.\cite{morton2009understanding,baia2006surface}  Various strategies have been developed to date for enhancing Raman signals, such as manipulating and tailoring the SERS substrates, modifying suitable experimental conditions for the measurements, and using noble metal nanoparticles for improvement.\cite{bar2021silicon,canamares2004surface,zeng2021zno,langer2019present} There are numerous challenges associated with achieving homogeneous hot spot distribution and reproducibility on SERS substrates, as signals are influenced by the distribution of analytes and nanoprobes, among many other factors. \cite{perez2020surface,langer2019present,wu2014high}

To achieve the ultrasensitivity of SERS detection of analytes, sophisticated surfaces with controlled morphology of plasmonic nanoparticles are required. \cite{de2011breaking,jin2005correlating,fan2014graphene} Efficient sample pretreatment methods have been adopted for extracting and pre\textendash concentrating analytes. Generally, the abundance of impurities within a sample can be several orders of magnitude higher than that of the analyte. As a result, these impurities facilitate non\textendash specific adsorption onto the SERS nanostructures and hinder the adsorption of the analytes, leading to inaccuracies in quantitative analysis during SERS measurements. Sample pretreatment that resolves the above problems enriches the analyte molecules for enhanced response signals during SERS analysis. \cite{perez2020surface,panneerselvam2018surface} Sample pretreatment demands selectivity and sensitivity for tracing target chemical species in complex media like food, serum, body fluids, and other biomolecules down to the level of fingerprints. \cite{perez2020surface}

Evaporation of a colloidal drop  is a simple approach for sample pretreatment that generates nanoparticle clusters with localized and abundant target molecules. Highly dense hot spots can be created in a small area to enhance SERS signals of concentrated analytes absorbed on the nanoparticles. Nevertheless, the undesired coffee\textendash stain effect in drop evaporation hinders the reproducibility and sensitivity of the approach due to the loss of constituents through the uncontrolled material deposition.\cite{mampallil2018review,yang2016ultrasensitive} However, Yang et al.\cite{yang2016ultrasensitive} carried out SERS detection by the evaporation of a single liquid droplet on the SLIP substrate. The lubricant layer is pre-applied on an entire surface of a structured substrate. Recent research showed that evaporation of a single  ouzo drop of ternary liquid mixture favours unpinned drop for solute concentrating during selective evaporation of the volatile co\textendash solvent.\cite{tan2016evaporation,tan2019porous} The preferential evaporation of the co\textendash solvent from the drop leads to over\textendash saturation of oil and spontaneously forms tiny oil droplets resulting from the Ouzo effect.\cite{tan2016evaporation,vitale2003liquid} These tiny droplets further coalesce and merge to form an oil ring around the rim of the drop and act as a lubricant for the aqueous drop shrinkage.\cite{tan2016evaporation,tan2019porous}  
Evaporation of self\textendash lubricating ouzo solution (a homogeneous ternary oil\textendash water\textendash ethanol solution) confers a promising technique for preventing the coffee ring effect.\cite{deegan1997capillary}  In the case of a colloidal droplet, the oil ring prevents the pinning of the contact line and enables the formation of a supraparticle.\cite{tan2019porous,tan2016evaporation,thayyil2021particle} 

Inspired by the aforementioned self\textendash lubricating phenomenon, in this work, we demonstrate an ultrasensitive approach by combining evaporation\textendash induced supraparticles (i.e., large clusters of nanoparticles ) and concentrating analytes in a self\textendash lubricating drop. The drop shrinks isotropically on a smooth substrate lubricated by an oil ring around the drop rim. The components in the drop include the nanoparticles and analytes from a deposit after the ﬁnal stage of evaporation. The limit of detection of the model compound rhodamine 6G is found to be $10^{-16} M$, close to a single molecule in the initial drop. There are major differences in our approaches compared to literature.\cite{yang2016ultrasensitive,zhang2019hydrophobic} Firstly, the drop is a ternary liquid and undergoes phase separation in evaporation. Secondly. the lubrication mechanism is from the oil layer formed by the evaporating drop itself. Lastly, the surface of the substrate is smooth and hydrophobic.  Our approach relies on an entirely new mechanism. Furthermore, we also demonstrate the ultrasensitive detection of a micro\textendash pollutant  in water and caffeine in human saliva from drinking tea. This work may serve as a new simple, and effective approach for sample preparation that can potentially be used for SERS detection in pharmaceuticals, food, environmental applications, and many other liquid samples. 

\section{Experimental Section}
\subsection {Chemicals and Materials}

Ethanol (90 $\%$) and 1\textendash octanol (95 $\%$) are the co\textendash solvents and oil phase in the ouzo drops, respectively. The water was obtained from a Milli\textendash Q water purification unit (Millipore Corporation, USA). For the synthesis of Ag nanoparticles, silver nitrate solution (0.1 N) and sodium citrate (99 $\%$) were purchased from Fisher Scientific (Canada). Potassium bromide (KBr) (99 $\%$, FT\textendash IR grade) and sodium borohydride (99 $\%$) were purchased from Sigma Aldrich (Canada). Cover glass (Fisher Scientific) and silicon wafer (University wafer, USA) were hydrophobized with a layer of octadecyl\textendash trichlorosilane (OTS, 98.9 $\%$) by following the protocol reported in our previous work.\cite{xu2017collective} The model analytes were rhodamine 6G (R6G) (95 $\%$) and triclosan (TC) powder acquired from TCI chemicals. Red Label Tea Powder (from India) and Nescafe coffee powder (from the US) were used to detect caffeine in saliva.

\subsection{Synthesis of Ag nanoparticles}

Ag nanoparticles (NPs) in variable sizes were synthesized by a seed\textendash mediated method reported by Frank et al. (refer to experimental section).\cite{frank2010synthesis} 1.25 $\times$ $10^{-2}$ M (200 mL) of sodium citrate, 3.75 $\times$ $10^{-4}$ M (500 mL) of silver nitrate, 5.0 $\times$ $10^{-2}$ M (500 mL) of hydrogen peroxide, 1.0 $\times$ $10^{-3}$ M (212.5 mL) of potassium bromide, and 5.0 $\times$ $10^{-3}$ M (250 mL) of sodium borohydride were prepared as a stock solution in amber glass bottles. Sequentially, all the reagents within the glass vial were added in accordance with Table S1. Afterward, a cap was placed on the vial, and the vial was carefully swirled to mix completely. Immediately, the progression of the reaction evolves, evident by the visual changes in the color of the mixture consistent with the growth of Ag NPs, corresponding to the KBr concentration in the vial Figure S1, in Supporting Information). 

The suspension solution of Ag NPs is prepared from 40 $\mu$L of synthesized nanoparticles base solution. The NPs mixture was centrifuged in 2 mL vials at 15,000 \textit{rpm}. The resulting NPs were identified at the bottom of the vial. The supernatant was discarded carefully using a pipette after completion of the centrifuge and re\textendash filled with water for the next round of centrifuge. The cycle was repeated twice to remove impurities from the NPs solution. Subsequently, after the centrifugal process, the supernatant was discarded, and the vial was further filled with 100 $\mu$L water with Ag NPs at the bottom (Figure S2). The Ag NPs suspension volume was further adjusted using water in accordance with the ternary mixture solution. Prior to the usage in ternary mixture preparation, the Ag suspension was sonicated for 20 min. 

The dispersity of Ag NPs suspension was well controlled by the selected volume of the initial Ag NPs synthesized solution used for centrifuging in a single vial. This procedure used 2 mL, 4 mL, 8 mL, and  12 mL of Ag NPs solutions to prepare initial volume\textendash based Ag NPs suspension. After forming Ag NPs suspension, different analyte solutions were mixed to obtain analyte/Ag suspension prior to further mixing with octanol/ethanol binary solution.

\subsection{Characterization of Ag nanoparticles}

UV Vis spectroscopy  (GENESYS 150) was used to obtain the absorbance spectra and size range of Ag NPs in the intial synthesized solution.  Field Emission Scanning Electron Microscope with Energy\textendash dispersive X\textendash  ray spectroscopy (Zeiss Sigma FESEM with EDX \& EBSD) was used to characterize the for morphology and elemental analysis. 
\begin{figure}[htp]
\centering
\includegraphics[width=\textwidth,height=\textheight,keepaspectratio]{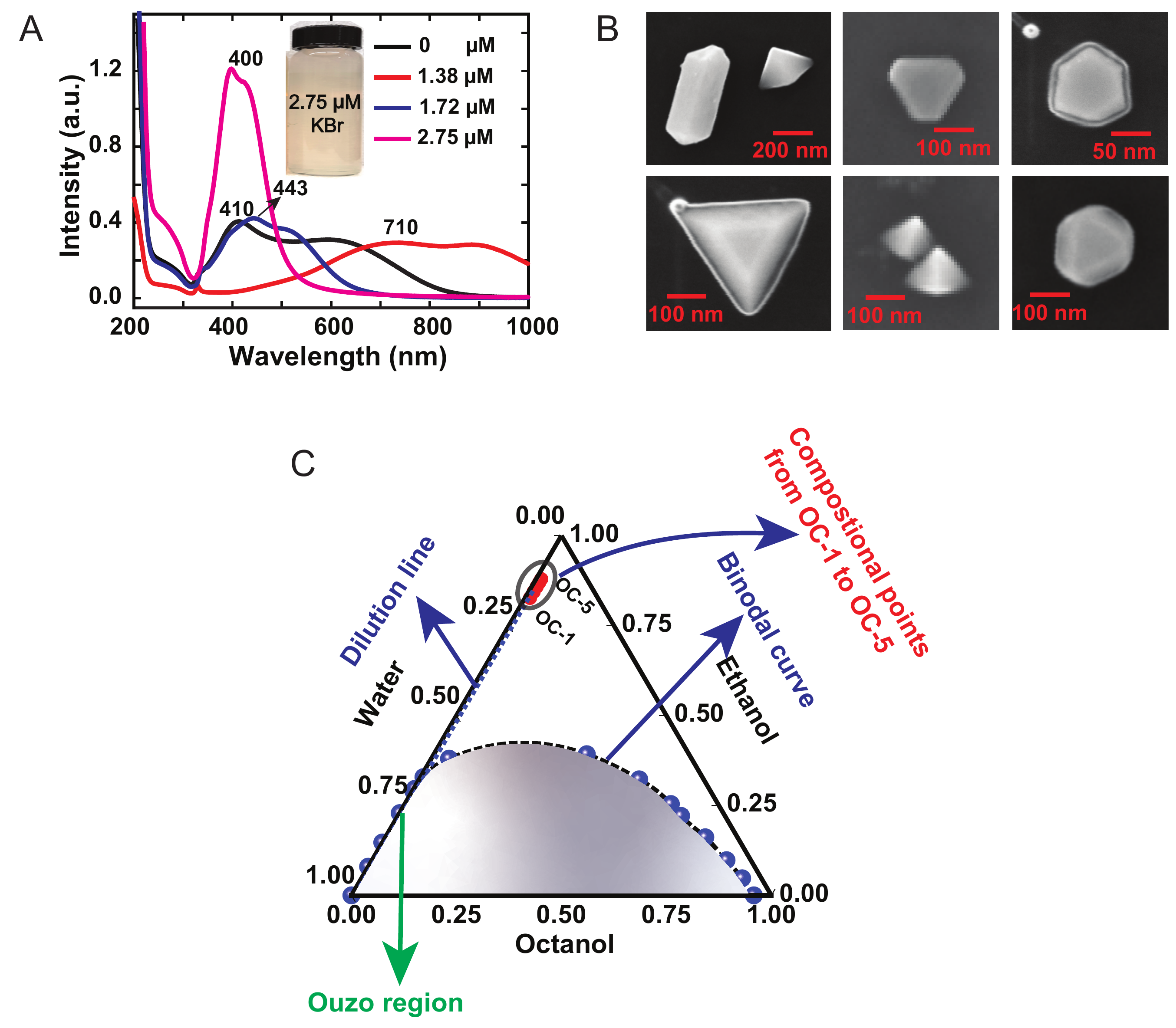}
\caption{(A) UV\textendash Vis absorbance spectra of Ag NPs synthesized at various KBr concentration, and (B) FESEM images of Ag NPs obtained at 2.75 $\mu$M KBr concentration. (C) Ternary phase diagram for octanol/ethanol/ water mixture with dilution line and all five corresponding mixture component ratios.}
\label{Figure 1}
\end{figure}

Figure S1 shows the color of the Ag NPs solution according to the amount of KBr addition.\cite{frank2010synthesis} UV\textendash Vis absorption spectral changes were also observed with the variation in the volume of KBr solution as shown in Figure \ref{Figure 1}a. The nanoparticles became smaller with an increase in the concentration of bromide addition from 0 $\mu$M to 2.75 $\mu$M and shifts the absorbance maxima from 710 nm to 400 nm as shown by UV\textendash Vis absorbance spectrum. Clearly, sharp absorption features were observed for 2.75 $\mu$M of KBr, whereas absorption features broadened with a lowered amount of KBr. This can be attributed to the limited growth of the Ag NPs due to the presence of KBr and the decrease in the sizes of nanoparticles owing to the strong bonds of bromide over the silver surface.\cite{frank2010synthesis,cathcart2009silver} Furthermore, the FESEM images in Figure \ref{Figure 1}b shows clearly that the Ag NPs synthesized by 2.75 $\mu$M of KBr comprises polyhedron type morphology with variable sizes.  

\subsection{Preparation of colloidal ouzo solutions}

The colloidal solution contained mixture of 82.45 \% (wt/wt) ethanol, 0.72 \% (wt/wt) octanol, and 16.83 \% (wt/wt) Ag NPs suspension. In the mixture system, octanol is termed the oil phase. The octanol/ethanol binary solution was initially prepared and sonicated for 20 min. A solution of 0.5 mL of Ag NP suspension was prepared by adding 0.4 mL of water directly into the Ag NP suspension vial (shown in Figure S2). The Ag NPs suspension was also sonicated for 20 min separately and then added into octanol/ethanol binary solution to prepare the final ternary ouzo solution. The ternary mixture was sonicated for a further 30 min before drop evaporation. The phase\textendash separation point was obtained from the octanol/ethanol/water ternary phase diagram as shown in Figure \ref{Figure 1}C. As explained in our previous work, the ratio of octanol in ouzo solutions is based on the integrated area between the dilution path and the binodal curve in the ouzo region.\cite{li2018formation} An ouzo region is determined by plotting the ternary solution composition values listed in Table S2 and the ternary diagram shown in Figure \ref{Figure 1}C. The dilution line represents the phase\textendash wise separation and sequential enrichment from ethanol to oil phase during the evaporation procedure. In the case of the dilution, the line enters into the shaded region, the ouzo effect comes into effect, and the microbubble starts to form inside the drop.

\subsubsection{Preparation of the analyte in Ag NPs suspension}

The R6G stock solutions at the concentrations of $10^{-3} M$ to $10^{-16} M$ were prepared by a titration method. 0.4 mL R6G stock solution was used to obtain R6G/Ag NPs suspension (Figure S3 A). 100 mL of $10^{-3} M$  ethanol\textendash triclosan stock solution was first prepared and used for the titration to make $10^{-4} M$, $10^{-5} M$, and $10^{-6} M$ of triclosan\textendash water solution. Triclosan\textendash aqueous solution, 0.4 mL was added to make triclosan/Ag NPs suspension. The analyte/Ag NPs suspensions were sonicated for 30 min before mixing with octanol/ethanol binary solution.

To further investigate, one of the authors collected blank saliva samples in amber glass bottles in the early morning between 5:30 to 6:30 AM after 12 h of overnight fasting. Caffeine\textendash containing saliva samples were collected just after the subject had taking tea or coffee. For example, one teaspoon of tea/coffee powder was used to make 50 mL of tea or coffee. Saliva was added directly into the Ag NPs suspension vial without further treatment or processing. In particular, 30 $\mu$L, 50 $\mu$L, 100 $\mu$L, 200 $\mu$L, and 300 $\mu$L of saliva samples were added to 1 mL of Ag NPs suspension vials to maintain the viscosity of Ag NPs suspension solution (Figure S3 B). Finally, before mixing the octanol/ethanol binary solution with saliva/Ag NPs, all suspensions were sonicated for 30 min.

\subsubsection{Preparation and analysis of analyte/Ag supraparticle  }  

Octadecyl\textendash trichlorosilane (OTS) coated silicon or cover glass was used as the substrate. Before the experiment, the substrate was first sonicated in ethanol for 30 min and dried at room temperature. The contact angle of the OTS\textendash coated substrate was measured as 105$^{\circ}$ by using a contact angle meter (Kruss DSC 100). Octanol/ethanol binary solution and analyte\textendash Ag suspension were mixed in appropriate weight ratios to obtain ternary ouzo solution and sonicated for 20 min to homogenize the three components in the ternary solution.

\begin{figure}[htp]
\includegraphics[width=0.95\textwidth,height=0.95\textheight,keepaspectratio]{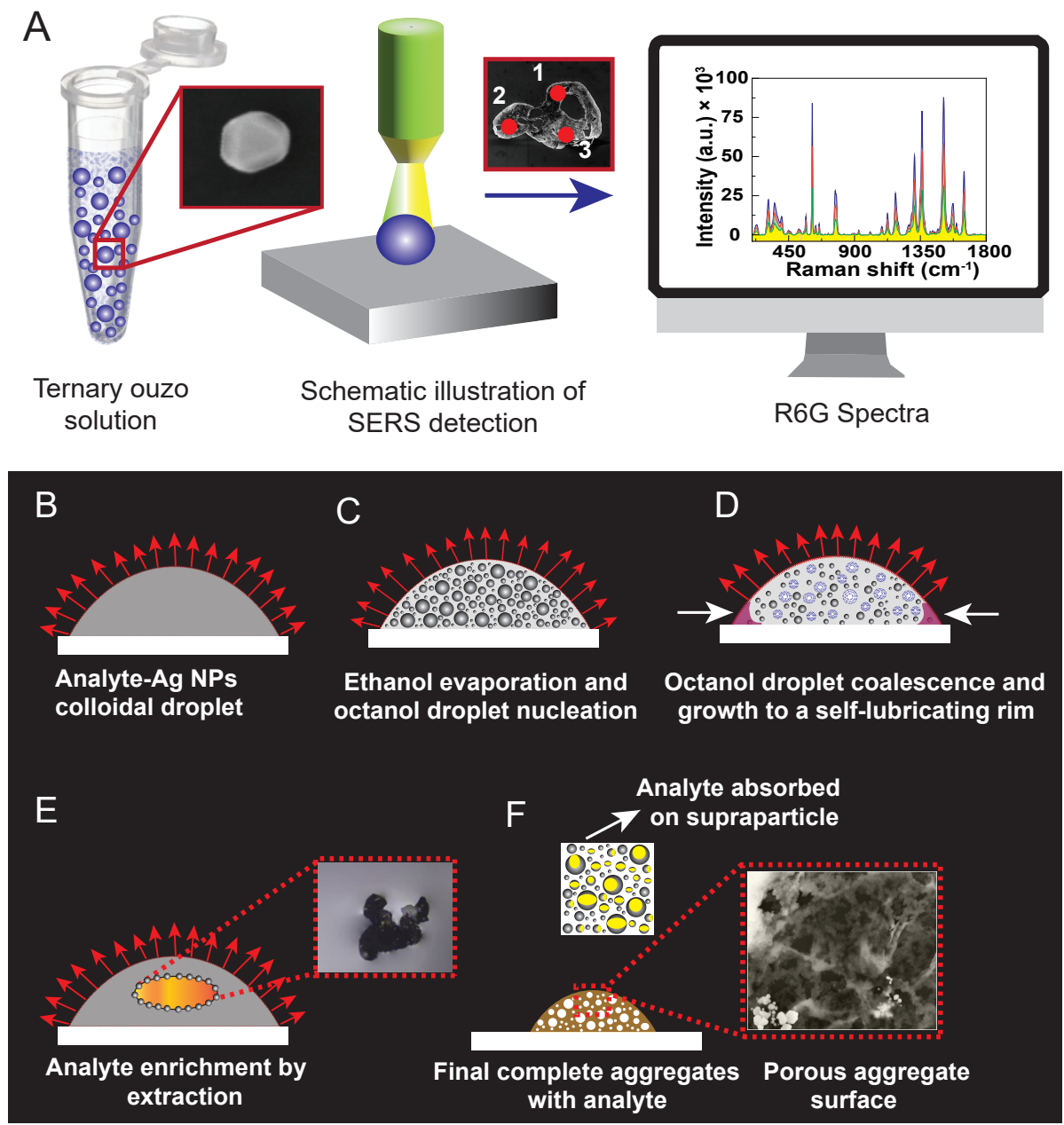}
\caption{(A) Schematic representation of SERS detection of the analyte using ouzo solution. A schematic representation  of porous Ag supraparticles from a self\textendash lubricating drop\textendash evaporation. (B) The initial state of analyte\textendash Ag NPs colloidal ouzo drop on a substrate, (C) Ethanol evaporation and resulting in the formation of octanol droplet nucleation, and (D). The oil ring's octanol droplet coalescence and shrinkage sweep nanoparticles from the substrate. (E) Analyte enrichment stage by the extraction process, and (F) final complete porous supraparticles on the surface.}
\label{figure4m}
\end{figure}

A schematic representation of the analyte detection procedure using silver supraparticles was described in Figure \ref{figure4m}A. A ternary solution was prepared by mixing octanol/ethanol binary solution with Ag analyte suspension. After 20 minutes of sonication, 2 mL of ouzo solution droplet was deposited on the substrate and allowed to evaporate. The supraparticles were formed using the ouzo method on OTS\textendash coated silicon substrates. In order to obtain the SERS spectra, three spots were chosen randomly on the surface of the supraparticles using a confocal Raman spectroscopy instrument. In our case, the evaporation time was varied from 30 min to 12 h. Optical images of the supraparticles were obtained using an upright optical microscope (NIKON H600l) coupled with a 10x and 100x lens.

Note that analyte/Ag supraparticles were used directly for SERS detection after 30 min and 12 h of evaporation (shown in Figure \ref{figure4m} B\textendash F). The SERS measurements were carried out on randomly selected spots on the surface of the supraparticles, and the corresponding intensity spectra were recorded. SERS measurements were carried out using a confocal Raman microscope (Renishaw). In particular, the detection was performed at 533 nm wavelength laser for R6G and at 785 nm for triclosan and saliva detection. The spectral scan range was selected for all the analytes from 350 nm\textendash 1800 nm range within objective 50x. For each measurement, 30 acquisitions were carried out with laser power of 0.1 \% for R6G and 0.5 \% for triclosan and saliva samples. The beam exposure time was selected as 15 s for each measurement. The same procedure was followed for the saliva samples and model compounds. Saliva or caffeine saliva was mixed with Ag NPs and other components of a ternary solution. The ouzo solution droplet (i.e., 2 $\mu$L) was used to form supraparticles on the substrate used directly for SERS detection after 30 min\textendash 60 min of evaporation. The colloidal Ouzo droplet evaporation process and 3D hot spots of SERS Illustration as shown in Figure \ref{figure4m} B\textendash F. The significant peaks with the respective bands for R6G, triclosan, saliva proteins, and caffeine are listed in supporting information Table S3 \textendash S6. Our approach has the advantage of evaporating drop with self-lubrication, which reduces the RAMAN sensitivity to the substrate properties. An established protocol was followed to prepare the OTS-coated Si substrate in the present study. The advancing and receding contact angles of water on our OTS-Si are 100\textdegree and 90\textdegree. The small hysteresis indicates a homogeneous substrate. In this procedure, we use the process of sedentary drop on the self-lubrication layer during the late stage of evaporation. Due to the hydrophobicity of the substrate, the self-lubrication effect dominates the drop shrinkage. As expected, the phenomenon is the same as what was observed on the OTS-Si substrate. The concentration of analytes and stepwise extraction stages come into existence without further modification of the surface morphology and topography. During optimization, both OTS-glass and OTS-Si substrates were used to study evaporation and SERS detection. The OTS-glass with Ag supraparticle has shown the least SERS activity due to transparency. Whereas OTS-Si with Ag supraparticle is shown excellent activity to SERS detection

\section{Results and discussion}

\subsection{Morphology and composition of supraparticles}

\begin{figure}[htp]
\centering
\includegraphics[width=\textwidth,height=\textheight,keepaspectratio]{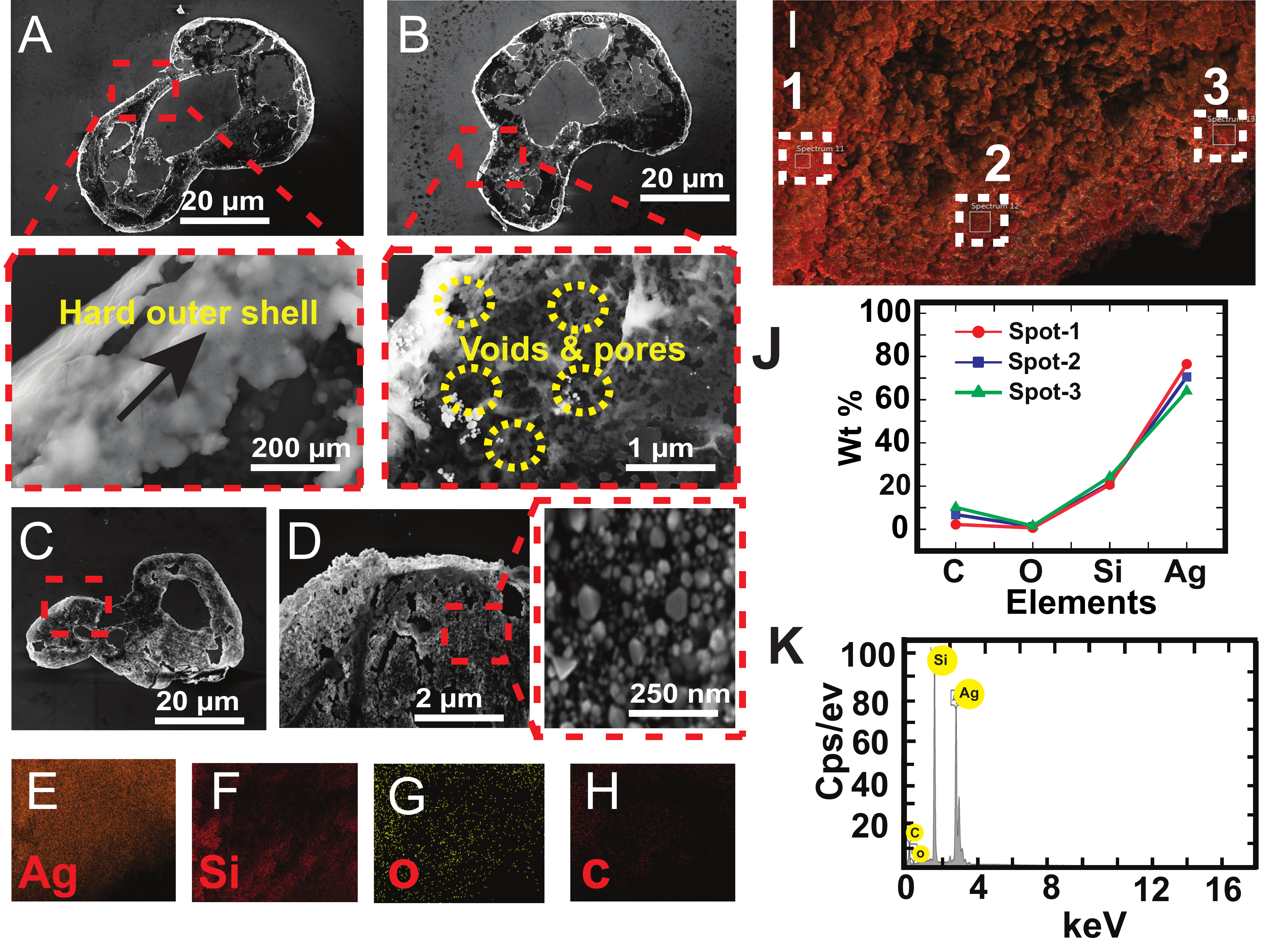}
\caption{FESEM images of (A) and (B) final porous aggregate morphology with hard exterior shell with porous surface structure. (C) Surface features of supraparticle and (D) Magnified image reveals sharp edges on the surface of the nanoparticles. (E\textendash H) Elemental mapping of supraparticle surface finds Ag, Si, O, and C. (I) EDX spectra images showing elements 1, 2, and 3. (J) Elemental quantification in composition percentage. (K) EDX analysis from a selected spot.}
\label{figure 4 latest}
\end{figure}

The supraparticles formed after drop evaporation vary in their shapes, influenced by various interfaces during the formation stage. The FESEM images in Figure \ref{figure 4 latest} A and B of final supraparticles show that the inner surface of the supraparticles is porous. The porosity may be due to voids from isolated residual water droplets (Figure S4 A\textendash E). Interestingly, the outer shell of the final supraparticles is cast firmly and rigidly.  The surface morphology reveals a sharp\textendash edged texture (Figure \ref{figure 4 latest} C and D). The shape of the supraparticles is relatively flat, possibly due to their formation near the air\textendash water interface.\cite{thayyil2021particle} Several kinds of the morphology of the Ag supraparticles are shown in Figure S5 A\textendash D. 

EDX spectra were recorded on the aggregate surface for a collective elemental information and distribution analysis using the element mapping technique. The element mapping shows an abundance of Ag atoms and Si atoms from the substrate (Figure \ref{figure 4 latest} E and F). The relatively low presence of O  atoms and C atoms detected in Figure \ref{figure 4 latest} G and H) could be associated with impurities. Figure \ref{figure 4 latest} I display the EDX analysis of three randomly selected points, indicating the dominance of the Ag atom and further proving the Ag\textendash rich aggregate surface. Elemental mapping analysis also confirmed the high percentages of Ag (i.e., 62 $\%$, 73 $\%$, and 79 $\%$) relative to other elements are shown in Figure \ref{figure 4 latest} J. A representative EDX spectrum of a point with corresponding elemental peaks is given in Figure \ref{figure 4 latest} K.




\subsection {Extremely low limit of detection (LOD)}

\begin{figure}[htp]
\centering
\includegraphics[width=\textwidth,height=\textheight,keepaspectratio]{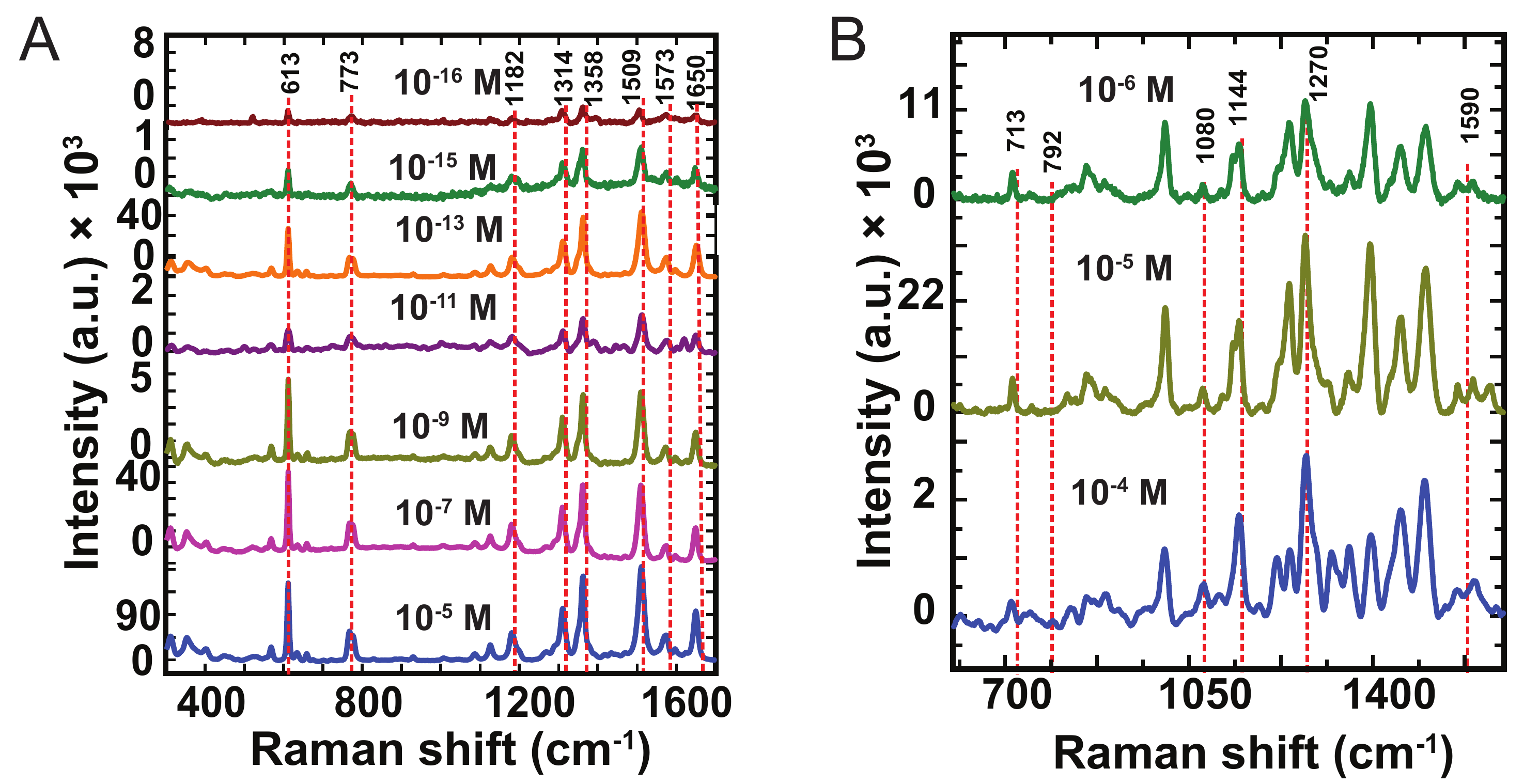}
\caption{After drying for 12 h, the limit of detection for supraparticles was achieved, (A) SERS spectra of R6G detection. (B) SERS spectra of triclosan detection.}
\label{figure 6A}
\end{figure}

Figure \ref{figure 6A} A shows that the LOD of R6G is as low as $10^{-16} M$ and the SERS spectra obtained during the optimization at $10^{-16} M$ are also depicted in Figure S8-S10. LOD level is comparable to other SERS approaches without any surface modification or pre\textendash treatment  requirement.\cite{yang2016ultrasensitive,zeng2021zno} Figure \ref{figure 6A} B shows the SERS spectra of triclosan with LOD of $10^{-6} M$. The relative detection level of triclosan is comparatively lower than that of R6G. The reason for the difference may be the partition coefficient (log $K_{0-W}$) (2.69 for R6G and 4.8 for triclosan \cite{li2019automated,olaniyan2016triclosan}) that directly affects the solubility and distribution of these two analytes in the octanol\textendash water biphasic system.\cite{chmiel2019impact} The level of LOD employed is comparable to other plasmonic nanoparticle labeled SERS approaches without any surface modification or pre\textendash treatment.\cite{mondal2020xenobiotic} All the spectral peak intensities are clearly visible and matched with the other SERS spectra. However, the appearance of few noise peaks in these SERS spectra is possibly due to detection by metal nanoparticles as a SERS active medium. As a result, the noise may come from very diverse sources, such as solution concentration caused by evaporation, metal nanoparticles aggregation at a particular unit area, the laser heating of the solution during a time of exposure, self-generated Raman spectrum of the metal nanoparticles, the SERS probes themselves, the noise from the optical system, spectrometer and data acquisition system, and noise from the operators.\cite{zhao2010background} SERS spectra were obtained by 6 random spots on the supraparticle surface to achieve LOD (Figure S6). The multiple detections at LOD further confirm the high reproducibility with a very minimal quantity of analyte solution.

\begin{figure}[htp]
\centering
\includegraphics[width=\textwidth,height=\textheight,keepaspectratio]{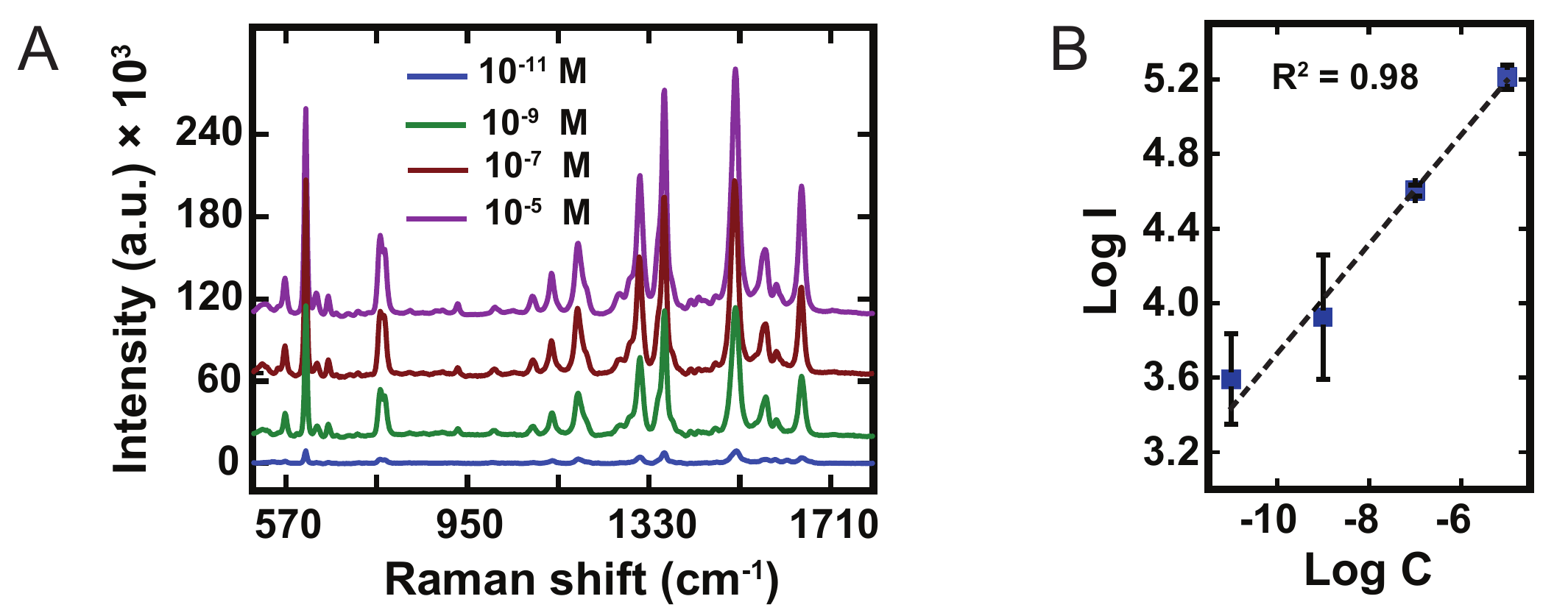}
\caption{ After 12 h of evaporation, (A) SERS spectra. (B) SERS intensity at 1508 $cm^{-1}$  as a function of R6G concentration.}
\label{figure 6}
\end{figure}

Three random spots of 1 $\mu$m diameter were selected for quantitative analysis of SERS on the surface of the supraparticles. At this particular concentration, the SERS intensities from three independent measurements are highly reproducible with minimal variations. The quantitative range is found from $10^{-5} M$ to $10^{-11} M$ (Figure \ref{figure 6} A). By integrating the peak intensity at 1509 $cm^{-1}$ in the spectrum, a linear relationship was obtained from a log\textendash log plot between the transformation of peak intensity and R6G concentration as illustrated in Figure \ref{figure 6} B. It is expressed by an empirical equation, $log \hspace{2pt}I = 0.292\hspace{4pt} log\hspace{4pt}C-6.65$, with good linear fitting variation coefficient ($R^{2}$ = 0.98). Where $C$ denotes R6G molar concentration, and $I$ is the SERS intensity. These results show that this method is suitable for the large\textendash range quantitative determination of SERS analysis. The repeated SERS spectra of R6G detection during 12 h of evaporation time for quantification are demonstrated in Figure S14.

\begin{figure}
\centering
\includegraphics[width=0.68\textwidth,height=0.68\textheight,keepaspectratio]{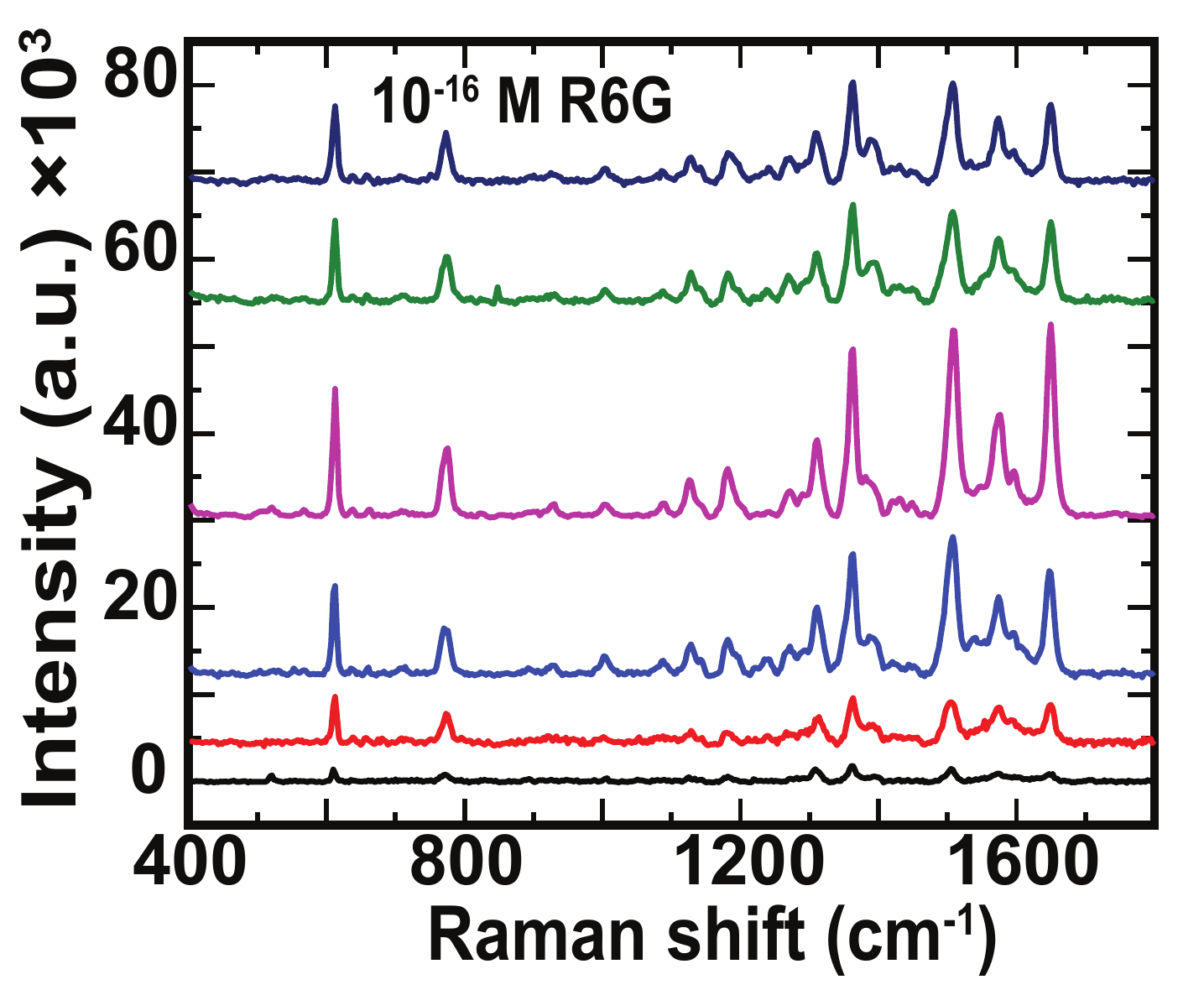}
\caption{SERS spectra detection at six spots over the supraparticles for $10^{-16}$ M concentration of R6G and 12 h drying.}
\label{fig:S8}

\end{figure}
The detection was initiated from the $10^{-5} M$ concentration set as the highest for the SERS measurement in our experiment, diluting it to $10^{-5} M$. It is important to note that the LOD was achieved by 6 random SERS spectra on an aggregate surface efficiently without requiring any fast scan rates or areal mapping over the selected area. The LOD of $10^{-16} M$ is also significant because very few molecules are present within the laser spot area. Multiple detections at low LOD further confirm the high reproducibility even with a very minimal quantity of analyte solution, as shown in Figure \ref{fig:S8}. The repeated SERS spectra of R6G detection during 12 h of evaporation time were also shown in Figures S8 and S9.\\

Analytes must be either in contact or close to noble metal nanoparticles to achieve optimal enhancement in this method of ouzo drop evaporation and self\textendash lubrication on a hydrophobic surface. The complexes accumulate at the apex of the droplet during evaporation, and the AgNP\textendash analyte interaction becomes intensified in a small zone after drying. The drying process brings together all the species within the droplet, improving the spectral quality of SERS measurements.  Below we show the effects of evaporation conditions on LOD and quantification by SERS.

\subsubsection {Effect of drop drying time and Ag NPs dispersity on SERS intensity}
\begin{figure}[htp]
\centering
\includegraphics[width=\textwidth,height=\textheight,keepaspectratio]{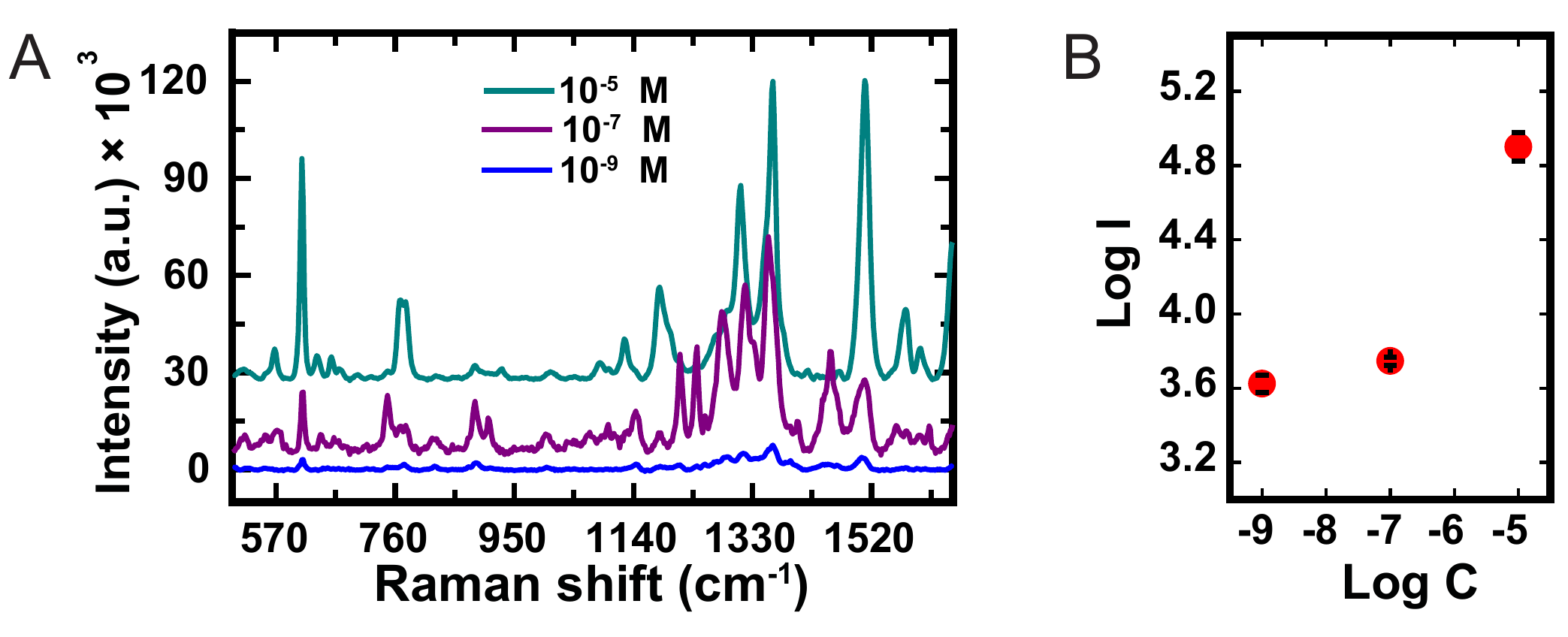}
\caption{  After 30 min of evaporation, (A) SERS spectra. (B) SERS intensity at 1508 $cm^{-1}$ as a function of R6G concentration.}
\label{figure 6_1}
\end{figure}

\begin{figure}[!htp]
\centering
\includegraphics[width=0.5\textwidth,height=0.5\textheight,keepaspectratio]{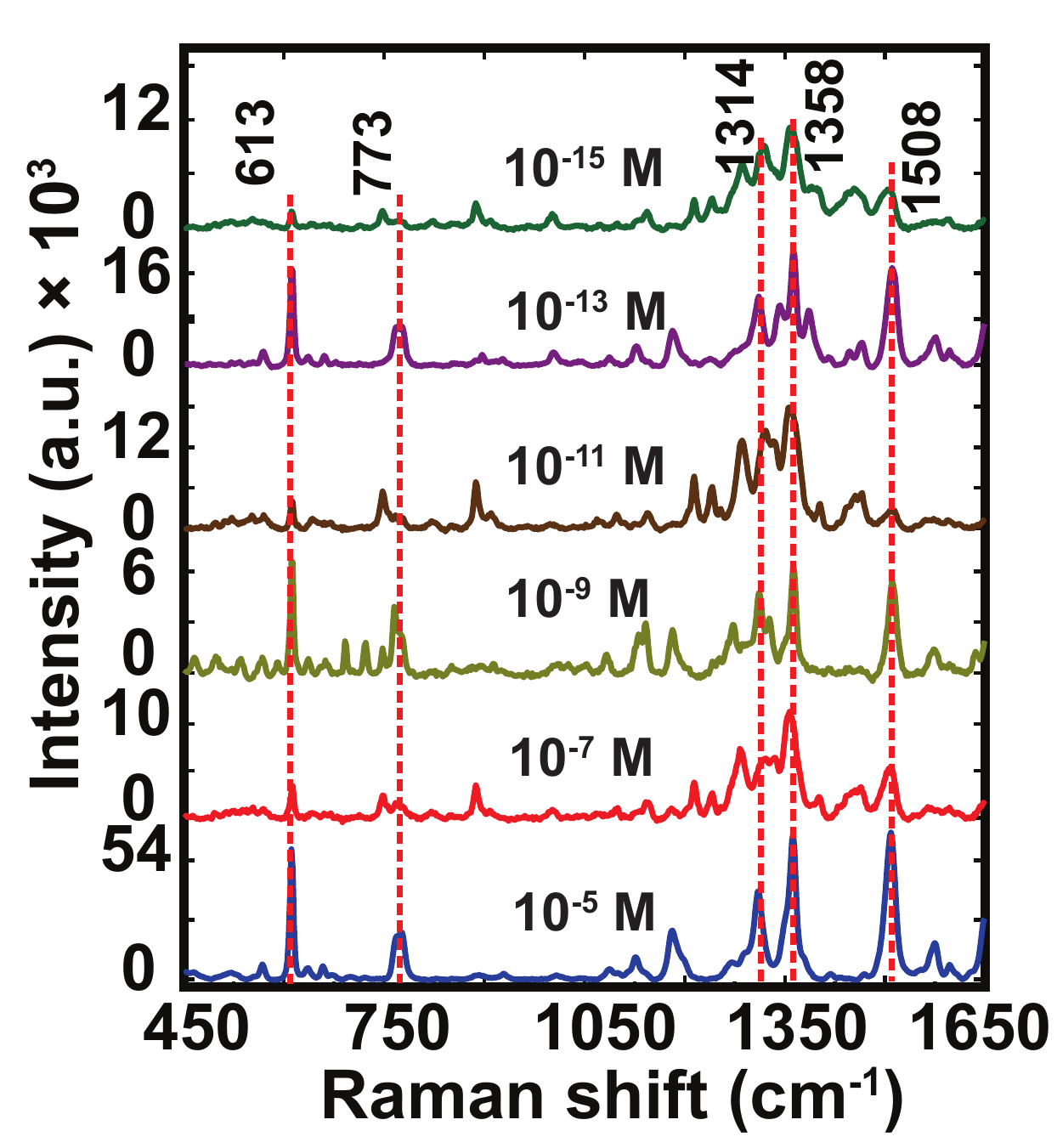}
\caption{SERS spectra of R6G detection after 30 min of drying time.}
\label{fig:S11}
\end{figure}

Figure \ref{figure 6_1} A shows the direct influence of drying time over the quantification of R6G SERS measurement. After 30 min of drying the sample, the linear relationship was calculated from $10^{-5}$ to $10^{-9}$ M as shown in Figure \ref{figure 6_1} B, which is comparatively less than that of SERS obtained after 12 h of drying time. All the other SERS peaks with different intensities are the function of a number of single Ag particles present on the supraparticle surface. Generally, supraparticles contribute to the majority of the total measured SERS signals. These strong continuum emission peaks only appear in conjunction with the R6G Raman signal; when the Raman signal blinks off, so does the continuum. Conversely, when the Raman signal reappears, the continuum does as well.\cite{michaels1999surface} Similarly, the LOD of the R6G is also relatively higher ($10^{-15}$ M) for 30 min sample drying process than for 12 h process (Figure \ref{fig:S11}). The repeated SERS spectra of R6G detection during 30 min of evaporation time are evidenced in Figure S7. These results further confirm the effect of the drying time of the sample on the SERS activity of R6G, which may be due to the slow evaporation of octanol.

\begin{figure}[htp]
\centering
\includegraphics[width=\textwidth,height=\textheight,keepaspectratio]{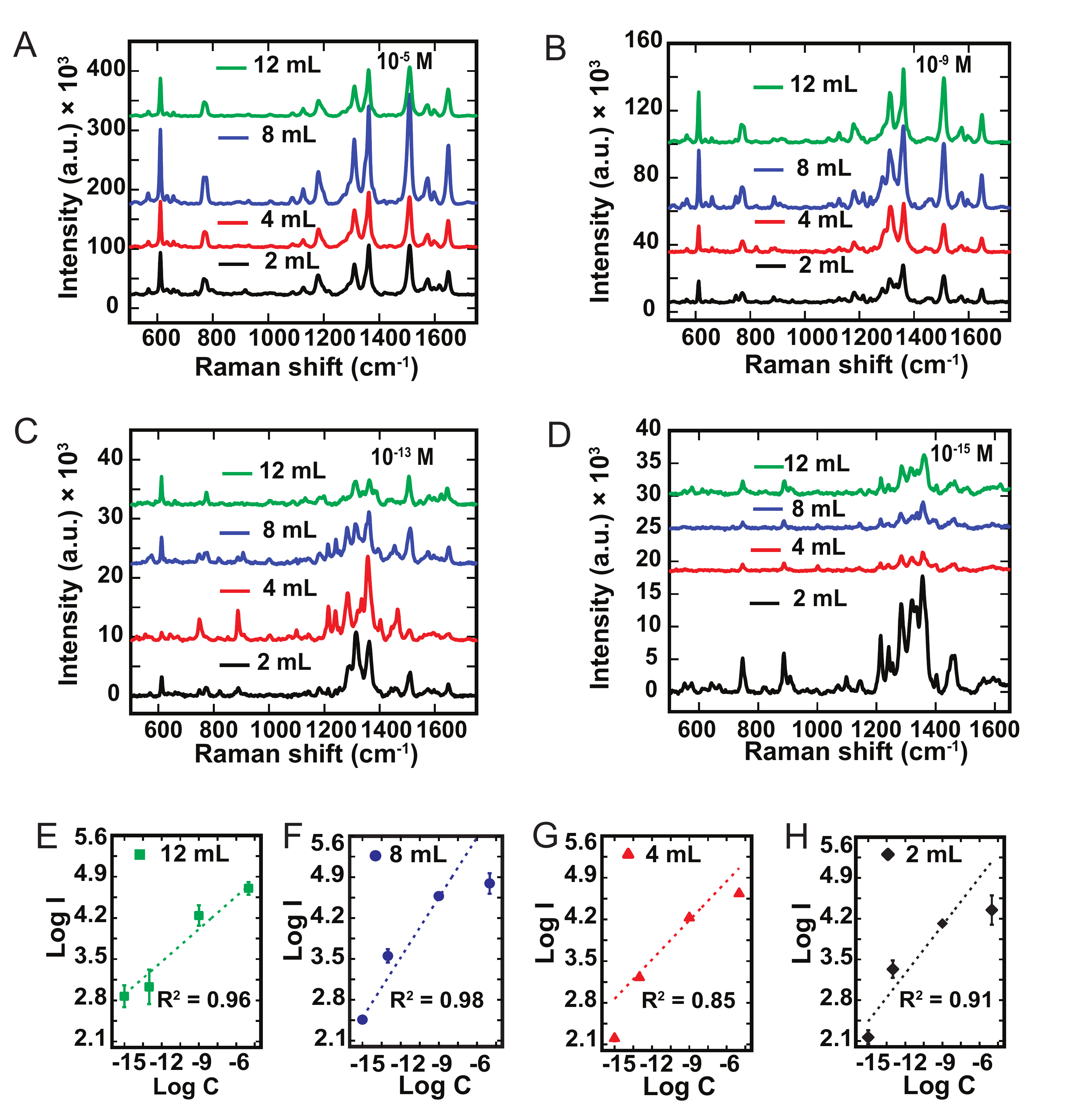}
\caption{ (A\textendash D) SERS intensity for detection of R6G using according to the Ag NPs volume dispersity. (E\textendash H) SERS intensity at 1508 $cm^{-1}$ as a function of R6G concentration as per Ag NPs dispersity.}
\label{fig:figure7}
\end{figure}

\begin{figure}[htp]
\centering
\includegraphics[width=0.9\textwidth,height=0.9\textheight,keepaspectratio]{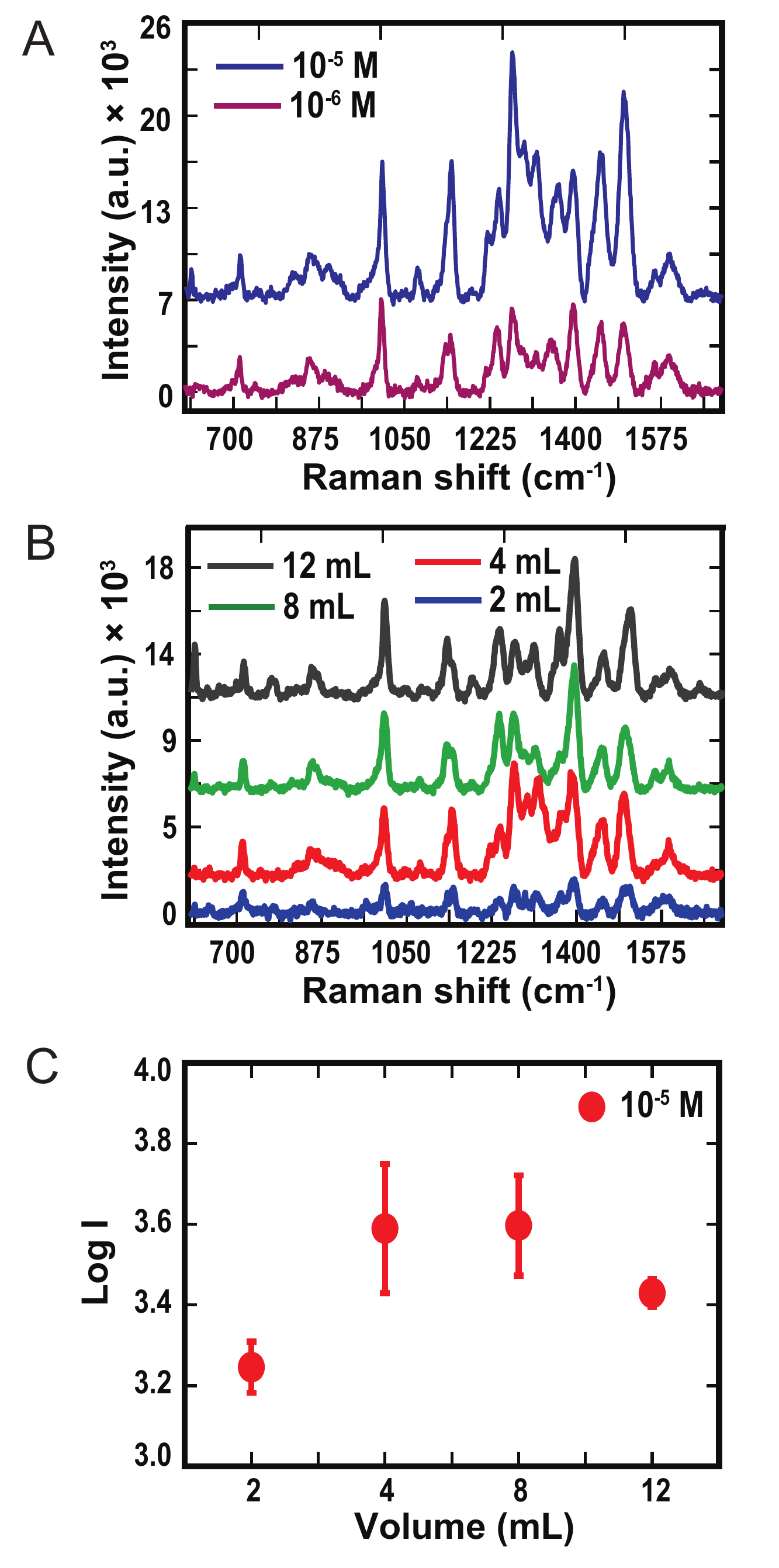}
\caption{ SERS spectra of triclosan, (A) at $10^{-5} M$ and $10^{-6} M$ concentrations after 30 min of evaporation process. (B) SERS spectra of triclosan at $10^{-5}$ M concentration as per Ag NPs volume dispersity. (C) Log\textendash intensity plot at 1508 $cm^{-1}$ as a function of Ag dispersity volume.}
\label{figure 8}
\end{figure}
The density of plasmonic nanostructures influences the intensity of SERS signals. The dispersity of Ag NPs in the ternary solution is directly associated with the density of supraparticles as the number of particles increases with the enhancement of NPs dispersion volume. To estimate the effect of Ag NPs dispersity on SERS detection ability of supraparticle, here we have controlled the initial Ag NPs concentration from 2 mL to 12 mL within the suspension preparation step as listed in Table S7\textendash S11. The Ag NPs suspension is added to the ternary analyte solution for obtaining different sizes of Ag/analyte supraparticles (as described in the experimental section above and shown in Figure S15). In our experiments, LOD of R6G obtained were up to $10^{-15} M$ for each volume of Ag NPs dispersed sample ( Figure\ref{fig:figure7} A\textendash D). 

At 2 mL of Ag NPs disperse sample, a linear relationship was obtained from a log\textendash log plot between transformation of the peak intensity and R6G concentration from $10^{-5} M$ to $10^{-15} M$ as shown in Figure \ref{fig:figure7} E. At 4 mL, 8 mL, and 12 mL Ag of dispersed samples, similar trends of quantification are also been observed (Figure \ref{fig:figure7} F\textendash H). In contrast, the quantification is independent of the  Ag NPs dispersity volumes for fixed concentration. For example, at $10^{-5} M$, the  log\textendash log plot  of peak intensity does not depend on R6G concentration  (shown in Figure S16). The enhancement factor (EF) for Ag supraparticle and without Ag supraparticle is quantified using the following relation:
\begin{equation*}
EF =   \frac{ I_{SERS} }{ I_{RAMAN} }   \frac{ N_{RAMAN} }{ N_{SERS} }  
\end{equation*}

Where, $I_{SERS}$ and $I_{RAMAN}$ are the intensity of in-plane vibrations of  $sp^{2}$ bonded carbon (C=C stretching vibrations (1580-1592) $cm^{-1}$ in the surface-enhanced spectrum and Raman spectrum. $N_{SERS}$ and $N_{RAMAN}$ are the concentration of analyte molecules sample for the SERS and Raman spectrum, respectively. In the case of Triclosan, the enhancement factor is in the range of $10^{-2}$ M, which is $~$2.9 times larger than simple drop evaporation without using any supraparticle. SERS can detect triclosan LOD irrespective of drying time, possibly due to the high concentration (Figure \ref{figure 8} A). The SERS spectra of triclosan with Ag Supraparticles at $10^{-2}$ M and Raman spectra  of triclosan are shown in Figure S10. The dependence of detection quantification on the dispersed Ag NPs volume is shown by using triclosan at $10^{-5} M$ in four different Ag NPs disperse ternary solutions (Figure \ref{figure 8} B and C). The peak intensity varies with Ag NPs disperse volume samples. The disparity obtained in SERS measurement could be due to variation in Ag NPs distribution on the aggregate surface, which may indicate the variation of density of plasmonic Ag NPs per unit area on the surface.
As explained by the above-mentioned phenomenon, this method is flexible in detecting LODs and quantification during SERS measurements by using dispersed Ag NPs. 

\subsubsection{Detection of caffeine from a small volume of saliva} 

The ability to detect target analytes with high specificity and sensitivity in any fluid is vital to analytical science and technology. Notably, detecting low abundance molecules in body fluids is necessary for the early identification of various diseases. Generally, all the body fluids for illicit drug use and diagnostic media for early disease detection. \cite{feng2015surface} Saliva can be collected conveniently, no\textendash destructively, and repeatedly in a non\textendash invasive way without any discomfort or pain for diagnosis purposes, unlike other body fluids such as blood. However, saliva is associated with many kinds of proteins that directly resemble many human disease transformations.\cite{liu2012saliva} For example, nucleic acids present in saliva are already used as tumor biomarkers for diagnosis.\cite{bonne2012salivary} 
The precise and rapid identification of metabolite and psychoactive substances prevents adverse health issues in accidental overconsumption.\cite{velivcka2021detection}

\begin{figure}[htp]
\centering
\includegraphics[width=\textwidth,height=\textheight,keepaspectratio]{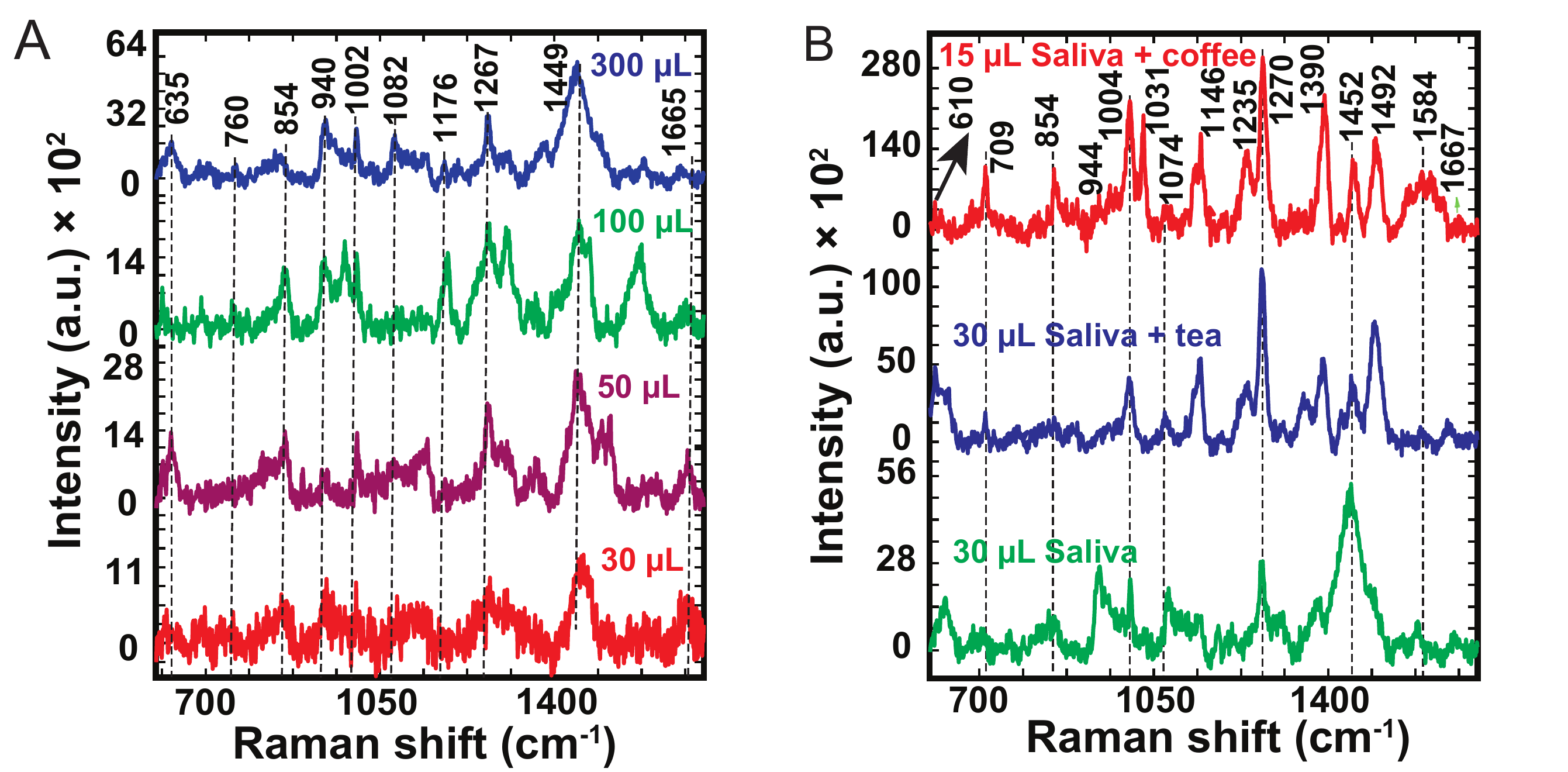}
\caption{SERS spectra of saliva by using supraparticle. (A) The volume of saliva varied from 300 $\mu$L to 30 $\mu$L after 1 h of sample preparation, and (B) caffeine SERS spectra in 30 $\mu$L of tea\textendash saliva and 15 $\mu$L of coffee\textendash saliva after 1 h of sample preparation.}
\label{figure 9}
\end{figure}

In this regard, we adopt the self\textendash lubricating drop for rapid analytical detection and identification of these psychoactive molecules. During the detection procedure, the SERS intensity is obtained from randomly selected spots on the surface of supraparticles. The signals of saliva were analysed in different volumes of saliva added in Ag NPs suspension. Figure \ref{figure 9} A shows the SERS spectra obtained from 30 $\mu$L, 50 $\mu$L, 100 $\mu$L, and 300 $\mu$L volume (Table S11) of saliva added Ag NPs suspension. Three corresponding trials at 30 $\mu$L of saliva detection using supraparticles are carried out, and SERS spectra are given in Figure S11. The main saliva proteins and other biomolecules are detected at 30 $\mu$L saliva addition. Tentative assignments of main saliva protein SERS bands to specific vibrational modes and biomolecules are listed in Table S5. \\

Especially, SERS peaks at 940 $cm^{-1}$, 1002 $cm^{-1}$, 1267 $cm^{-1}$, and 1449 $cm^{-1}$ were prominently detected which are generally found in a healthy individual.\cite{feng2015surface} However, these peak intensities were increased with the quantity of saliva addition, which further confirms the rapid\textendash non\textendash invasive and highly sensitive SERS detection of saliva chemical components by this method within 30 min. Gonchukov et al.\cite{gonchukov2011raman} reported the drop drying method. There were strong peaks at 1653 and 1002 $cm^{-1}$ due to the polypeptide backbone of a protein because of amide and aromatic ring breathing, respectively. The presence of CH stretching peak at 1444 $cm^{-1}$ attributed to glycoproteins in mucin matrices. They depicted Raman bands at 852 and 1128 $cm^{-1}$ correspond to C-N stretching and $CH_{3}$ rocking, C-O vibrations without determining any biomolecule identity. Additionally, they found that a Raman shift of 750 cm is related to oxygenated hemocyanin because of the vibration of the O-O stretching. The repeated SERS spectra of saliva  detection during 30 min of evaporation time are given shown in Figure S11. 

We investigated saliva samples after drinking coffee and tea to detect caffeine using SERS. After 1 h of complete evaporation of all the liquid phases, caffeine molecules were detected using SERS on the surface of the supraparticle. The minimum amount of coffee and tea saliva samples for SERS detection are 15 $\mu$L and 30 $\mu$L. 
SERS spectra in Figure \ref{figure 9} B,  shows that SERS peaks at 1235 $cm^{-1}$ and 1390 $cm^{-1}$ are associated with C\textendash N stretching and symmetric stretching in caffeine molecules. The repeated SERS spectra of caffeine in tea saliva and coffee saliva during 1 h of evaporation time are shown in Figures S12 and S13. As a result, the peaks appear distinctly and confirm the caffeine detection in low\textendash volume samples. All the positions and assignments of SERS peaks of caffeine molecules are listed in Table S6. SERS peak at 1492 $cm^{-1}$ predominantly in tea and coffee samples is associated with C=C deformation in phenyl rings. It is worth noting that the appearance of 1031 $cm^{-1}$ SERS peak in only coffee samples is associated with C\textendash C twisting mode in phenylalanine present in saliva samples. SERS peaks at 709 $cm^{-1}$, 854 $cm^{-1}$, 1004 $cm^{-1}$, 1270 $cm^{-1}$, 1452 $cm^{-1}$, and 1575 $cm^{-1}$ are signatures of caffeine molecules in saliva samples.

\subsection{Mechanisms of the self\textendash lubricating drop for ultrasensitive SERS detection}

\begin{figure}[htp]
\centering
\includegraphics[width=\textwidth,height=\textheight,keepaspectratio]{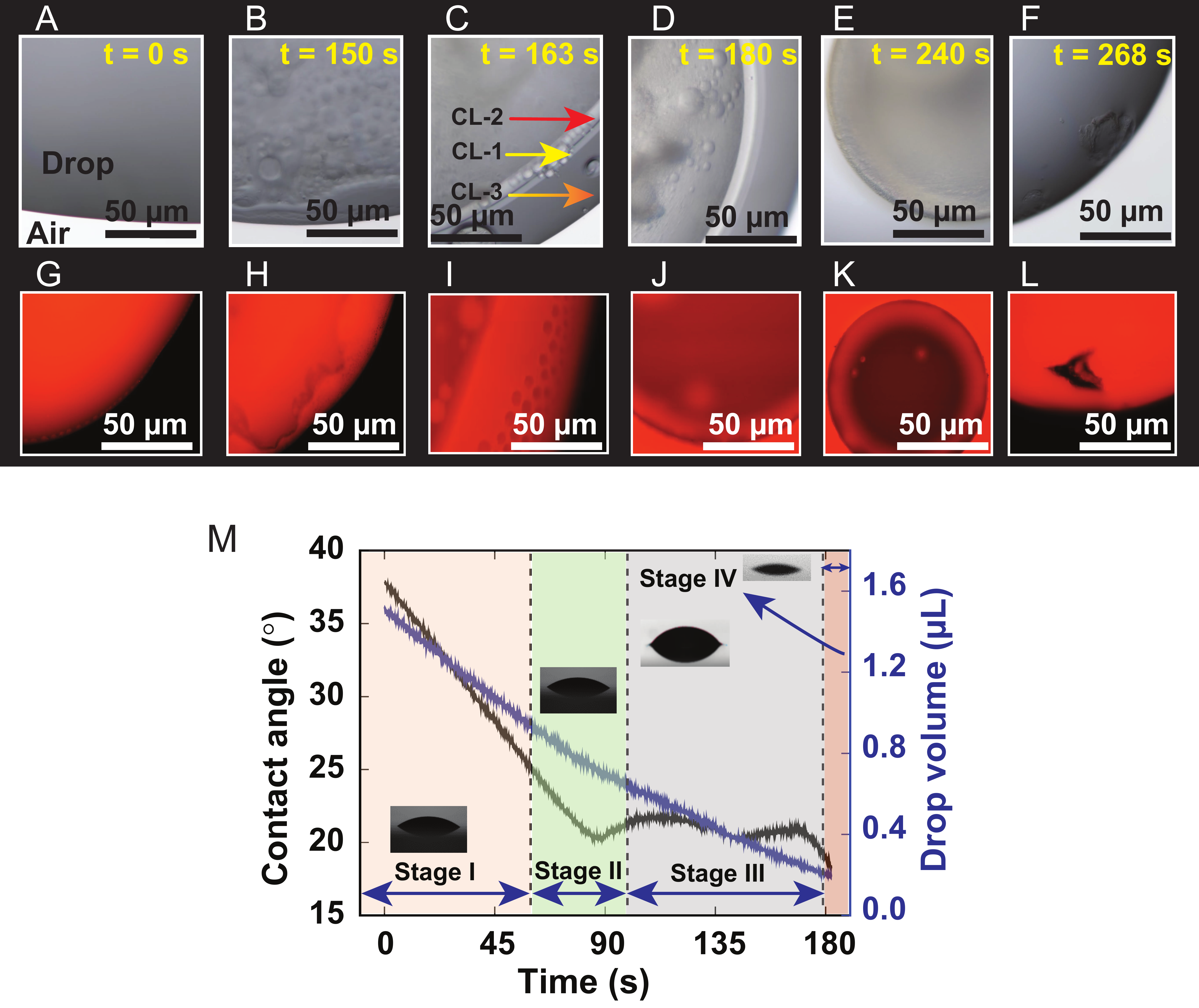}
\caption{Evaporation of octanal (oil phase)/ethanol/Ag suspension ternary solution,
	(A) at the early stage, the ouzo drop is transparent, (B) microdroplets nucleate at the rim of the drop after 150 $s$, and (C) the microdroplets are convected throughout the whole ouzo drop, and oil\textendash oversaturation leads to oil droplet growth and coalescence. As the water drop evaporates further, (D) and (E) most of the oil droplets coalesce into an oil ring at the rim of the drop with Ag NPs making it more turbid. (F) When water evaporation increases, Ag NPs accumulate in the bulk of the remaining spherical\textendash cap\textendash shaped sessile oil drop. Similarly, each stage is described illustrating R6G dye during evaporation (G\textendash L). (M) Contact angle and droplet volume variation with time during droplet evaporation and various stages of ouzo solution evaporation.}
\label{figure 3}
\end{figure}

Optical images of evaporating ouzo droplet are captured to illustrate the extraction, pore formation, and self\textendash lubrication process, as shown in Figure \ref{figure 3} A\textendash L. The mechanism for self\textendash lubrication is explained by the following four stages of ouzo drop evaporation \cite{tan2019porous,tan2017self,koshkina2021surface}:
In stage 1, the ouzo drop appears as a transparent spherical cap just after the drop generation since the components are uniformly distributed within the droplet as shown in Figure \ref{figure 3} A and G. The evaporation process lowers the ethanol concentration inside the drop, and the mixture becomes an oil over\textendash saturated. However, after 150 s at the onset of stage 2, oil microdroplets form at the rim and then distribute throughout the whole drop. This mechanism is mainly attributed to the convective flow inside the drop (Figure \ref{figure 3} B and H). It is important to note that the dispersed droplets may act as the sites for nanoparticle aggregation. \cite{tan2019porous,koos2011capillary} The oil ring appears from the deposition of coalesced oil droplets around the drop boundary at 163 s. Figure \ref{figure 3} C and I reveal the presence of three types of contact lines (CL) near the oil ring; CL\textendash 1, where the mixture, surface, and oil meet. CL\textendash 2, where the mixture, oil, and air meet. CL\textendash 3, where oil, substrate, and air meet. At stage 3, most of the oil droplets have coalesced into an oil ring at the rim of the drop at 180 s (Figure \ref{figure 3} D and J).

The exterior portion of the oil ring in drop becomes transparent. Subsequently, nanoparticles begin to get attached to the oil microdroplets and concentrate within the ring (at t = 240 s in Figure \ref{figure 3} E and N). At a later stage, water evaporates from the drop as the oil ring lubricates the shrinking drop process. The oil ring is then forced to slide inwards and reduce circumstantially to a minimum, which leads to supraparticles sculpting into a three\textendash dimensional structure. The shrinkage of the oil ring causes the levitation of supraparticles. In stage 4, due to the dominance of capillary force in a small oil drop, supraparticles float, as observed for other supraparticles. \cite{tan2019porous,koshkina2021surface} The supraparticles float over the surface of residual oil drop represented in Figure \ref{figure 3}F and L. The change in  drop volume  and  contact angle $\theta$ during the evaporation process are recorded in Figure \ref{figure 3}M. 


In stage\textendash 1 and stage\textendash 2, the drop volume and  contact angle decrease rapidly since ethanol evaporates rapidly. Self\textendash lubricating ring forms at the transition from stage\textendash 2 to stage\textendash 3. As a result of the lower evaporation rate of water, the slope of the drop volume line was reduced. The contact angle remained almost constant because of self\textendash lubrication over the substrate after evaporation of the water. In stage\textendash 4, a tiny oil drop remains on the surface of the substrate after supraparticle assembly.

Three main factors contribute to the very low concentration detection based on evaporation of self\textendash lubricating drop. First of all, the sequential evaporation of a more volatile liquid to a less volatile liquid over the hydrophobic substrate leads to enrichment of the analyte concentration within the final liquid phase. Secondly, the phase\textendash wise evaporation causes the extraction of analyte solute components from one phase to another. In coordination with evaporation and extraction, self\textendash lubrication further helps to reduce the loss of solvent remaining over the substrate surface. The lower LOD of R6G benefits from the high partition coefficient(log $K_{O-W}$) of R6G in octanol\textendash water, i.e., 2.69. \cite{lahnstein2008pulmonary} Similarly, for triclosan, the log $K_{O-W}$  is approximately 4.8.\cite{li2019automated} Accordingly, the concentration of these compounds in octanol is several orders of magnitude higher than that of water at equilibrium. 
In the final stage, solute extraction and transfer to the octanol phase follow the appearance of octanol micro\textendash droplets. The third factor to assist the high SERS intensity may be the final morphology and structure of the supraparticle. The large surface area of these pockets anchors all analytes to Ag NPs.


\cleardoublepage
\section{Conclusions}
In this work, we developed an ultrasensitive SERS detection process that overcomes long\textendash standing limitations of drop evaporation due to the coffee\textendash stain effect provides precise delivery of analytes to hot spots of chemical or biological species. Our results manifested a smooth hydrophobic surface that can be self\textendash lubricated by a colloidal ouzo drop to form porous supraparticles. The lowest LOD obtained for R6G and triclosan were of $10^{-16}$ M and $10^{-6}$ M respectively from an initial volume of  2 $\mu$L. The quantification of R6G was also achieved from  $10^{-5}$ M to $10^{-11}$ M after thoroughly drying the samples. The additional advantage is that the intensity of SERS detection is robust, not affected by the dispersity of the Ag NPs, which provides flexibility with respect to plasmonic nanoparticle preparation and detection. As a demonstration, we successfully detected caffeine present in saliva after consuming tea or coffee by our procedure. We presented the mechanism for the low limit of detection that is ascribed to the combined process of  extraction that analytes, droplet\textendash templated porous structure formation and self\textendash lubrication during  colloidal ouzo drop evaporation. This approach opens avenues for direct and highly sensitive chemical analysis and detection of aqueous environment  pollutants in medical, forensics, or illicit drug control.   

\section*{Credit author statement}
\textbf{Tulsi Satyavir Dabodiya:} Conducting experiment, Data curation, Formal Analysis, Investigation, Methodology, Writing\textendash Original Draft.  \textbf{Somasekhara Goud Sontti:} Project administration, visualization, Writing\textendash Original Draft, Writing\textendash review \& editing. \textbf{Zixiang Wei:} Data curation, Formal Analysis, Methodology, validation. \textbf{Qiuyun Lu:} Methodology.
\textbf{Romain Billet:} Data curation, Methodology. 
\textbf{Arumugam Vadivel Murugan:} Resources.  \textbf{Xuehua Zhang:} Conceptualization, Funding acquisition, Investigation, Methodology, Project administration, Resources, Supervision, Writing\textendash Original Draft, Writing\textendash review \& editing.
\cleardoublepage
\section*{Declaration of competing interest}
The authors declare that they have no known competing financial interests or personal relationships that could have appeared to influence the work reported in this paper.

\section*{Supporting Information}
supporting  Information  is  available.

\section{Acknowledgements}
The project is supported by Discovery project and Alliance Grant from the Natural Science and Engineering Research Council of Canada (NSERC), and by Advanced Program from Alberta Innovates. This research was undertaken, in part, thanks to funding from the Canada Research Chairs Program. Tulsi Satyavir Dabodiya acknowledges the financial support of the Overseas Visiting Doctoral Fellowship program (OVDF) from the Science and Engineering Research Board (SERB), SERB\textendash OVDF 2019\textendash 2020 program, Department of Science and Technology, Government of India. T.S Dabodiya thanks Dr. Esma Khatun  for sharing the SERS spectrum without evaporation. 
\bibliography{Literature}

\begin{thebibliography}{50}
\expandafter\ifx\csname natexlab\endcsname\relax\def\natexlab#1{#1}\fi
\providecommand{\url}[1]{\texttt{#1}}
\providecommand{\href}[2]{#2}
\providecommand{\path}[1]{#1}
\providecommand{\DOIprefix}{doi:}
\providecommand{\ArXivprefix}{arXiv:}
\providecommand{\URLprefix}{URL: }
\providecommand{\Pubmedprefix}{pmid:}
\providecommand{\doi}[1]{\href{http://dx.doi.org/#1}{\path{#1}}}
\providecommand{\Pubmed}[1]{\href{pmid:#1}{\path{#1}}}
\providecommand{\bibinfo}[2]{#2}
\ifx\xfnm\relax \def\xfnm[#1]{\unskip,\space#1}\fi
\bibitem[{Campion and Kambhampati(1998)}]{campion1998surface}
\bibinfo{author}{A.~Campion}, \bibinfo{author}{P.~Kambhampati},
\newblock \bibinfo{journal}{Chem. Soc. Rev.} \bibinfo{volume}{27}
  (\bibinfo{year}{1998}) \bibinfo{pages}{241--250}.
\bibitem[{Jarvis and Goodacre(2008)}]{jarvis2008characterisation}
\bibinfo{author}{R.~M. Jarvis}, \bibinfo{author}{R.~Goodacre},
\newblock \bibinfo{journal}{Chem. Soc. Rev.} \bibinfo{volume}{37}
  (\bibinfo{year}{2008}) \bibinfo{pages}{931--936}.
\bibitem[{Qian and Nie(2008)}]{qian2008single}
\bibinfo{author}{X.-M. Qian}, \bibinfo{author}{S.~M. Nie},
\newblock \bibinfo{journal}{Chem. Soc. Rev.} \bibinfo{volume}{37}
  (\bibinfo{year}{2008}) \bibinfo{pages}{912--920}.
\bibitem[{Adu et~al.(2012)Adu, Williams, Reber, Jayasingha, Gutierrez, and
  Sumanasekera}]{adu2012probing}
\bibinfo{author}{K.~W. Adu}, \bibinfo{author}{M.~D. Williams},
  \bibinfo{author}{M.~Reber}, \bibinfo{author}{R.~Jayasingha},
  \bibinfo{author}{H.~R. Gutierrez}, \bibinfo{author}{G.~U. Sumanasekera},
\newblock \bibinfo{journal}{J. Nanotechnol.} \bibinfo{volume}{2012}
  (\bibinfo{year}{2012}).
\bibitem[{Perumal et~al.(2021)Perumal, Wang, Attia, Dinish, and
  Olivo}]{perumal2021towards}
\bibinfo{author}{J.~Perumal}, \bibinfo{author}{Y.~Wang},
  \bibinfo{author}{A.~B.~E. Attia}, \bibinfo{author}{U.~Dinish},
  \bibinfo{author}{M.~Olivo},
\newblock \bibinfo{journal}{Nanoscale} \bibinfo{volume}{13}
  (\bibinfo{year}{2021}) \bibinfo{pages}{553--580}.
\bibitem[{Bantz et~al.(2011)Bantz, Meyer, Wittenberg, Im, Kurtulu{\c{s}}, Lee,
  Lindquist, Oh, and Haynes}]{bantz2011recent}
\bibinfo{author}{K.~C. Bantz}, \bibinfo{author}{A.~F. Meyer},
  \bibinfo{author}{N.~J. Wittenberg}, \bibinfo{author}{H.~Im},
  \bibinfo{author}{{\"O}.~Kurtulu{\c{s}}}, \bibinfo{author}{S.~H. Lee},
  \bibinfo{author}{N.~C. Lindquist}, \bibinfo{author}{S.-H. Oh},
  \bibinfo{author}{C.~L. Haynes},
\newblock \bibinfo{journal}{Phys. Chem. Chem. Phys.} \bibinfo{volume}{13}
  (\bibinfo{year}{2011}) \bibinfo{pages}{11551--11567}.
\bibitem[{Fan et~al.(2020)Fan, Andrade, and Brolo}]{fan2020review}
\bibinfo{author}{M.~Fan}, \bibinfo{author}{G.~F. Andrade},
  \bibinfo{author}{A.~G. Brolo},
\newblock \bibinfo{journal}{Anal. Chim. Acta} \bibinfo{volume}{1097}
  (\bibinfo{year}{2020}) \bibinfo{pages}{1--29}.
\bibitem[{Huang et~al.(2020)Huang, Liu, Gong, Wu, Fan, Wang, and
  Brolo}]{huang2020detection}
\bibinfo{author}{Y.~Huang}, \bibinfo{author}{W.~Liu},
  \bibinfo{author}{Z.~Gong}, \bibinfo{author}{W.~Wu}, \bibinfo{author}{M.~Fan},
  \bibinfo{author}{D.~Wang}, \bibinfo{author}{A.~G. Brolo},
\newblock \bibinfo{journal}{ACS Sens.} \bibinfo{volume}{5}
  (\bibinfo{year}{2020}) \bibinfo{pages}{2933--2939}.
\bibitem[{Sylvia et~al.(2000)Sylvia, Janni, Klein, and
  Spencer}]{sylvia2000surface}
\bibinfo{author}{J.~M. Sylvia}, \bibinfo{author}{J.~A. Janni},
  \bibinfo{author}{J.~Klein}, \bibinfo{author}{K.~M. Spencer},
\newblock \bibinfo{journal}{Anal. Chem.} \bibinfo{volume}{72}
  (\bibinfo{year}{2000}) \bibinfo{pages}{5834--5840}.
\bibitem[{Bharati et~al.(2021)Bharati, Soma et~al.}]{bharati2021flexible}
\bibinfo{author}{M.~S.~S. Bharati}, \bibinfo{author}{V.~R. Soma}, et~al.,
\newblock \bibinfo{journal}{Opto-electron. Adv.} \bibinfo{volume}{4}
  (\bibinfo{year}{2021}) \bibinfo{pages}{210048}.
\bibitem[{Morton and Jensen(2009)}]{morton2009understanding}
\bibinfo{author}{S.~M. Morton}, \bibinfo{author}{L.~Jensen},
\newblock \bibinfo{journal}{J. Am. Chem. Soc.} \bibinfo{volume}{131}
  (\bibinfo{year}{2009}) \bibinfo{pages}{4090--4098}.
\bibitem[{Baia et~al.(2006)Baia, Baia, Astilean, and Popp}]{baia2006surface}
\bibinfo{author}{M.~Baia}, \bibinfo{author}{L.~Baia},
  \bibinfo{author}{S.~Astilean}, \bibinfo{author}{J.~Popp},
\newblock \bibinfo{journal}{Appl. Phys. Lett.} \bibinfo{volume}{88}
  (\bibinfo{year}{2006}) \bibinfo{pages}{143121}.
\bibitem[{Shiohara et~al.(2014)Shiohara, Wang, and
  Liz-Marzán}]{shiohara2020recent}
\bibinfo{author}{A.~Shiohara}, \bibinfo{author}{Y.~Wang},
  \bibinfo{author}{L.~M. Liz-Marzán},
\newblock \bibinfo{journal}{J. Photochem. Photobiol. C} \bibinfo{volume}{21}
  (\bibinfo{year}{2014}) \bibinfo{pages}{2--25}.
\bibitem[{Bar et~al.(2021)Bar, de~Barros, de~Camargo, Pereira, Merces, Shimizu,
  Sigoli, Bufon, and Mazali}]{bar2021silicon}
\bibinfo{author}{J.~Bar}, \bibinfo{author}{A.~de~Barros},
  \bibinfo{author}{D.~H. de~Camargo}, \bibinfo{author}{M.~P. Pereira},
  \bibinfo{author}{L.~Merces}, \bibinfo{author}{F.~M. Shimizu},
  \bibinfo{author}{F.~A. Sigoli}, \bibinfo{author}{C.~C.~B. Bufon},
  \bibinfo{author}{I.~O. Mazali},
\newblock \bibinfo{journal}{ACS Appl. Mater. Interfaces} \bibinfo{volume}{13}
  (\bibinfo{year}{2021}) \bibinfo{pages}{36482--36491}.
\bibitem[{Canamares et~al.(2004)Canamares, Garcia-Ramos, Domingo, and
  Sanchez-Cortes}]{canamares2004surface}
\bibinfo{author}{M.~Canamares}, \bibinfo{author}{J.~Garcia-Ramos},
  \bibinfo{author}{C.~Domingo}, \bibinfo{author}{S.~Sanchez-Cortes},
\newblock \bibinfo{journal}{J. Raman Spectrosc.} \bibinfo{volume}{35}
  (\bibinfo{year}{2004}) \bibinfo{pages}{921--927}.
\bibitem[{Zeng et~al.(2021)Zeng, Wang, Du, Liu, Wang, Xu, and
  Wang}]{zeng2021zno}
\bibinfo{author}{Y.~Zeng}, \bibinfo{author}{F.~Wang}, \bibinfo{author}{D.~Du},
  \bibinfo{author}{S.~Liu}, \bibinfo{author}{C.~Wang}, \bibinfo{author}{Z.~Xu},
  \bibinfo{author}{H.~Wang},
\newblock \bibinfo{journal}{Appl. Surf. Sci.} \bibinfo{volume}{544}
  (\bibinfo{year}{2021}) \bibinfo{pages}{148924}.
\bibitem[{Langer et~al.(2019)Langer, Jimenez~de Aberasturi, Aizpurua,
  Alvarez-Puebla, Augui{\'e}, Baumberg, Bazan, Bell, Boisen, Brolo
  et~al.}]{langer2019present}
\bibinfo{author}{J.~Langer}, \bibinfo{author}{D.~Jimenez~de Aberasturi},
  \bibinfo{author}{J.~Aizpurua}, \bibinfo{author}{R.~A. Alvarez-Puebla},
  \bibinfo{author}{B.~Augui{\'e}}, \bibinfo{author}{J.~J. Baumberg},
  \bibinfo{author}{G.~C. Bazan}, \bibinfo{author}{S.~E. Bell},
  \bibinfo{author}{A.~Boisen}, \bibinfo{author}{A.~G. Brolo}, et~al.,
\newblock \bibinfo{journal}{ACS nano} \bibinfo{volume}{14}
  (\bibinfo{year}{2019}) \bibinfo{pages}{28--117}.
\bibitem[{P{\'e}rez-Jim{\'e}nez et~al.(2020)P{\'e}rez-Jim{\'e}nez, Lyu, Lu,
  Liu, and Ren}]{perez2020surface}
\bibinfo{author}{A.~I. P{\'e}rez-Jim{\'e}nez}, \bibinfo{author}{D.~Lyu},
  \bibinfo{author}{Z.~Lu}, \bibinfo{author}{G.~Liu}, \bibinfo{author}{B.~Ren},
\newblock \bibinfo{journal}{Chem. Sci.} \bibinfo{volume}{11}
  (\bibinfo{year}{2020}) \bibinfo{pages}{4563--4577}.
\bibitem[{Wu et~al.(2014)Wu, Hang, Komadina, Ling, and Li}]{wu2014high}
\bibinfo{author}{Y.~Wu}, \bibinfo{author}{T.~Hang},
  \bibinfo{author}{J.~Komadina}, \bibinfo{author}{H.~Ling},
  \bibinfo{author}{M.~Li},
\newblock \bibinfo{journal}{Nanoscale} \bibinfo{volume}{6}
  (\bibinfo{year}{2014}) \bibinfo{pages}{9720--9726}.
\bibitem[{De~Angelis et~al.(2011)De~Angelis, Gentile, Mecarini, Das, Moretti,
  Candeloro, Coluccio, Cojoc, Accardo, Liberale et~al.}]{de2011breaking}
\bibinfo{author}{F.~De~Angelis}, \bibinfo{author}{F.~Gentile},
  \bibinfo{author}{F.~Mecarini}, \bibinfo{author}{G.~Das},
  \bibinfo{author}{M.~Moretti}, \bibinfo{author}{P.~Candeloro},
  \bibinfo{author}{M.~Coluccio}, \bibinfo{author}{G.~Cojoc},
  \bibinfo{author}{A.~Accardo}, \bibinfo{author}{C.~Liberale}, et~al.,
\newblock \bibinfo{journal}{Nat. Photonics} \bibinfo{volume}{5}
  (\bibinfo{year}{2011}) \bibinfo{pages}{682--687}.
\bibitem[{Jin et~al.(2005)Jin, Jureller, Kim, and Scherer}]{jin2005correlating}
\bibinfo{author}{R.~Jin}, \bibinfo{author}{J.~E. Jureller},
  \bibinfo{author}{H.~Y. Kim}, \bibinfo{author}{N.~F. Scherer},
\newblock \bibinfo{journal}{J. Am. Chem. Soc.} \bibinfo{volume}{127}
  (\bibinfo{year}{2005}) \bibinfo{pages}{12482--12483}.
\bibitem[{Fan et~al.(2014)Fan, Lee, Pedireddy, Zhang, Liu, and
  Ling}]{fan2014graphene}
\bibinfo{author}{W.~Fan}, \bibinfo{author}{Y.~H. Lee},
  \bibinfo{author}{S.~Pedireddy}, \bibinfo{author}{Q.~Zhang},
  \bibinfo{author}{T.~Liu}, \bibinfo{author}{X.~Y. Ling},
\newblock \bibinfo{journal}{Nanoscale} \bibinfo{volume}{6}
  (\bibinfo{year}{2014}) \bibinfo{pages}{4843--4851}.
\bibitem[{Panneerselvam et~al.(2018)Panneerselvam, Liu, Wang, Liu, Ding, Li,
  Wu, and Tian}]{panneerselvam2018surface}
\bibinfo{author}{R.~Panneerselvam}, \bibinfo{author}{G.-K. Liu},
  \bibinfo{author}{Y.-H. Wang}, \bibinfo{author}{J.-Y. Liu},
  \bibinfo{author}{S.-Y. Ding}, \bibinfo{author}{J.-F. Li},
  \bibinfo{author}{D.-Y. Wu}, \bibinfo{author}{Z.-Q. Tian},
\newblock \bibinfo{journal}{Chem. Commun.} \bibinfo{volume}{54}
  (\bibinfo{year}{2018}) \bibinfo{pages}{10--25}.
\bibitem[{Mampallil and Eral(2018)}]{mampallil2018review}
\bibinfo{author}{D.~Mampallil}, \bibinfo{author}{H.~B. Eral},
\newblock \bibinfo{journal}{Adv. Colloid Interface Sci.} \bibinfo{volume}{252}
  (\bibinfo{year}{2018}) \bibinfo{pages}{38--54}.
\bibitem[{Yang et~al.(2016)Yang, Dai, Stogin, and
  Wong}]{yang2016ultrasensitive}
\bibinfo{author}{S.~Yang}, \bibinfo{author}{X.~Dai}, \bibinfo{author}{B.~B.
  Stogin}, \bibinfo{author}{T.-S. Wong},
\newblock \bibinfo{journal}{Proc. Natl. Acad. Sci.} \bibinfo{volume}{113}
  (\bibinfo{year}{2016}) \bibinfo{pages}{268--273}.
\bibitem[{Tan et~al.(2016)Tan, Diddens, Lv, Kuerten, Zhang, and
  Lohse}]{tan2016evaporation}
\bibinfo{author}{H.~Tan}, \bibinfo{author}{C.~Diddens},
  \bibinfo{author}{P.~Lv}, \bibinfo{author}{J.~G. Kuerten},
  \bibinfo{author}{X.~Zhang}, \bibinfo{author}{D.~Lohse},
\newblock \bibinfo{journal}{Proc. Natl. Acad. Sci.} \bibinfo{volume}{113}
  (\bibinfo{year}{2016}) \bibinfo{pages}{8642--8647}.
\bibitem[{Tan et~al.(2019)Tan, Wooh, Butt, Zhang, and Lohse}]{tan2019porous}
\bibinfo{author}{H.~Tan}, \bibinfo{author}{S.~Wooh}, \bibinfo{author}{H.-J.
  Butt}, \bibinfo{author}{X.~Zhang}, \bibinfo{author}{D.~Lohse},
\newblock \bibinfo{journal}{Nat. Commun.} \bibinfo{volume}{10}
  (\bibinfo{year}{2019}) \bibinfo{pages}{1--8}.
\bibitem[{Vitale and Katz(2003)}]{vitale2003liquid}
\bibinfo{author}{S.~A. Vitale}, \bibinfo{author}{J.~L. Katz},
\newblock \bibinfo{journal}{Langmuir} \bibinfo{volume}{19}
  (\bibinfo{year}{2003}) \bibinfo{pages}{4105--4110}.
\bibitem[{Deegan et~al.(1997)Deegan, Bakajin, Dupont, Huber, Nagel, and
  Witten}]{deegan1997capillary}
\bibinfo{author}{R.~D. Deegan}, \bibinfo{author}{O.~Bakajin},
  \bibinfo{author}{T.~F. Dupont}, \bibinfo{author}{G.~Huber},
  \bibinfo{author}{S.~R. Nagel}, \bibinfo{author}{T.~A. Witten},
\newblock \bibinfo{journal}{Nature} \bibinfo{volume}{389}
  (\bibinfo{year}{1997}) \bibinfo{pages}{827--829}.
\bibitem[{Thayyil~Raju et~al.(2021)Thayyil~Raju, Koshkina, Tan, Riedinger,
  Landfester, Lohse, and Zhang}]{thayyil2021particle}
\bibinfo{author}{L.~Thayyil~Raju}, \bibinfo{author}{O.~Koshkina},
  \bibinfo{author}{H.~Tan}, \bibinfo{author}{A.~Riedinger},
  \bibinfo{author}{K.~Landfester}, \bibinfo{author}{D.~Lohse},
  \bibinfo{author}{X.~Zhang},
\newblock \bibinfo{journal}{ACS nano} \bibinfo{volume}{15}
  (\bibinfo{year}{2021}) \bibinfo{pages}{4256--4267}.
\bibitem[{Zhang et~al.(2019)Zhang, You, Yang, Gao, Jiang, and
  Yin}]{zhang2019hydrophobic}
\bibinfo{author}{C.~Zhang}, \bibinfo{author}{T.~You},
  \bibinfo{author}{N.~Yang}, \bibinfo{author}{Y.~Gao},
  \bibinfo{author}{L.~Jiang}, \bibinfo{author}{P.~Yin},
\newblock \bibinfo{journal}{Food Chem.} \bibinfo{volume}{287}
  (\bibinfo{year}{2019}) \bibinfo{pages}{363}.
\bibitem[{Xu et~al.(2017)Xu, Yu, Peng, Lu, Lei, Lohse, and
  Zhang}]{xu2017collective}
\bibinfo{author}{C.~Xu}, \bibinfo{author}{H.~Yu}, \bibinfo{author}{S.~Peng},
  \bibinfo{author}{Z.~Lu}, \bibinfo{author}{L.~Lei},
  \bibinfo{author}{D.~Lohse}, \bibinfo{author}{X.~Zhang},
\newblock \bibinfo{journal}{Soft Matter} \bibinfo{volume}{13}
  (\bibinfo{year}{2017}) \bibinfo{pages}{937--944}.
\bibitem[{Frank et~al.(2010)Frank, Cathcart, Maly, and
  Kitaev}]{frank2010synthesis}
\bibinfo{author}{A.~J. Frank}, \bibinfo{author}{N.~Cathcart},
  \bibinfo{author}{K.~E. Maly}, \bibinfo{author}{V.~Kitaev},
\newblock \bibinfo{journal}{J. Chem. Educ.} \bibinfo{volume}{87}
  (\bibinfo{year}{2010}) \bibinfo{pages}{1098--1101}.
\bibitem[{Cathcart et~al.(2009)Cathcart, Frank, and
  Kitaev}]{cathcart2009silver}
\bibinfo{author}{N.~Cathcart}, \bibinfo{author}{A.~J. Frank},
  \bibinfo{author}{V.~Kitaev},
\newblock \bibinfo{journal}{Chem. Commun.}  (\bibinfo{year}{2009})
  \bibinfo{pages}{7170--7172}.
\bibitem[{Li et~al.(2018)Li, Bao, Yu, and Zhang}]{li2018formation}
\bibinfo{author}{M.~Li}, \bibinfo{author}{L.~Bao}, \bibinfo{author}{H.~Yu},
  \bibinfo{author}{X.~Zhang},
\newblock \bibinfo{journal}{J. Phys. Chem. C} \bibinfo{volume}{122}
  (\bibinfo{year}{2018}) \bibinfo{pages}{8647--8654}.
\bibitem[{Li et~al.(2019)Li, Dyett, and Zhang}]{li2019automated}
\bibinfo{author}{M.~Li}, \bibinfo{author}{B.~Dyett},
  \bibinfo{author}{X.~Zhang},
\newblock \bibinfo{journal}{Anal. Chem.} \bibinfo{volume}{91}
  (\bibinfo{year}{2019}) \bibinfo{pages}{10371--10375}.
\bibitem[{Olaniyan et~al.(2016)Olaniyan, Mkwetshana, and
  Okoh}]{olaniyan2016triclosan}
\bibinfo{author}{L.~Olaniyan}, \bibinfo{author}{N.~Mkwetshana},
  \bibinfo{author}{A.~Okoh},
\newblock \bibinfo{journal}{Springerplus} \bibinfo{volume}{5}
  (\bibinfo{year}{2016}) \bibinfo{pages}{1--17}.
\bibitem[{Chmiel et~al.(2019)Chmiel, Mieszkowska, Kempi{\'n}ska-Kupczyk,
  Kot-Wasik, Namie{\'s}nik, and Mazerska}]{chmiel2019impact}
\bibinfo{author}{T.~Chmiel}, \bibinfo{author}{A.~Mieszkowska},
  \bibinfo{author}{D.~Kempi{\'n}ska-Kupczyk}, \bibinfo{author}{A.~Kot-Wasik},
  \bibinfo{author}{J.~Namie{\'s}nik}, \bibinfo{author}{Z.~Mazerska},
\newblock \bibinfo{journal}{Microchem. J.} \bibinfo{volume}{146}
  (\bibinfo{year}{2019}) \bibinfo{pages}{393--406}.
\bibitem[{Mondal and Subramaniam(2020)}]{mondal2020xenobiotic}
\bibinfo{author}{S.~Mondal}, \bibinfo{author}{C.~Subramaniam},
\newblock \bibinfo{journal}{ACS Sustain. Chem. Eng.} \bibinfo{volume}{8}
  (\bibinfo{year}{2020}) \bibinfo{pages}{7639--7648}.
\bibitem[{Zhao et~al.(2010)Zhao, Huang, Shi, Cao, Yang, Gu
  et~al.}]{zhao2010background}
\bibinfo{author}{W.~Zhao}, \bibinfo{author}{S.~Huang},
  \bibinfo{author}{C.~Shi}, \bibinfo{author}{L.~Cao},
  \bibinfo{author}{X.~Yang}, \bibinfo{author}{C.~Gu}, et~al.,
\newblock \bibinfo{journal}{J. Opt. Commun.} \bibinfo{volume}{31}
  (\bibinfo{year}{2010}) \bibinfo{pages}{88}.
\bibitem[{Michaels et~al.(1999)Michaels, Nirmal, and
  Brus}]{michaels1999surface}
\bibinfo{author}{A.~M. Michaels}, \bibinfo{author}{M.~Nirmal},
  \bibinfo{author}{L.~Brus},
\newblock \bibinfo{journal}{J. Am. Chem. Soc.} \bibinfo{volume}{121}
  (\bibinfo{year}{1999}) \bibinfo{pages}{9932}.
\bibitem[{Feng et~al.(2015)Feng, Huang, Lin, Chen, Xu, Li, Huang, Pan, Chen,
  and Zeng}]{feng2015surface}
\bibinfo{author}{S.~Feng}, \bibinfo{author}{S.~Huang},
  \bibinfo{author}{D.~Lin}, \bibinfo{author}{G.~Chen}, \bibinfo{author}{Y.~Xu},
  \bibinfo{author}{Y.~Li}, \bibinfo{author}{Z.~Huang},
  \bibinfo{author}{J.~Pan}, \bibinfo{author}{R.~Chen},
  \bibinfo{author}{H.~Zeng},
\newblock \bibinfo{journal}{Int. J. Nanomed.} \bibinfo{volume}{10}
  (\bibinfo{year}{2015}) \bibinfo{pages}{537}.
\bibitem[{Liu and Duan(2012)}]{liu2012saliva}
\bibinfo{author}{J.~Liu}, \bibinfo{author}{Y.~Duan},
\newblock \bibinfo{journal}{Oral Oncol.} \bibinfo{volume}{48}
  (\bibinfo{year}{2012}) \bibinfo{pages}{569--577}.
\bibitem[{Bonne and Wong(2012)}]{bonne2012salivary}
\bibinfo{author}{N.~J. Bonne}, \bibinfo{author}{D.~T. Wong},
\newblock \bibinfo{journal}{Genome Med.} \bibinfo{volume}{4}
  (\bibinfo{year}{2012}) \bibinfo{pages}{1--12}.
\bibitem[{Veli{\v{c}}ka et~al.(2021)Veli{\v{c}}ka, Zacharovas,
  Adomavi{\v{c}}i{\=u}t{\.e}, and {\v{S}}ablinskas}]{velivcka2021detection}
\bibinfo{author}{M.~Veli{\v{c}}ka}, \bibinfo{author}{E.~Zacharovas},
  \bibinfo{author}{S.~Adomavi{\v{c}}i{\=u}t{\.e}},
  \bibinfo{author}{V.~{\v{S}}ablinskas},
\newblock \bibinfo{journal}{Spectrochim. Acta A Mol. Biomol. Spectrosc.}
  \bibinfo{volume}{246} (\bibinfo{year}{2021}) \bibinfo{pages}{118956}.
\bibitem[{Gonchukov et~al.(2011)Gonchukov, Sukhinina, Bakhmutov, and
  Minaeva}]{gonchukov2011raman}
\bibinfo{author}{S.~Gonchukov}, \bibinfo{author}{A.~Sukhinina},
  \bibinfo{author}{D.~Bakhmutov}, \bibinfo{author}{S.~Minaeva},
\newblock \bibinfo{journal}{Laser Phys. Lett.} \bibinfo{volume}{9}
  (\bibinfo{year}{2011}) \bibinfo{pages}{73}.
\bibitem[{Tan et~al.(2017)Tan, Diddens, Versluis, Butt, Lohse, and
  Zhang}]{tan2017self}
\bibinfo{author}{H.~Tan}, \bibinfo{author}{C.~Diddens},
  \bibinfo{author}{M.~Versluis}, \bibinfo{author}{H.-J. Butt},
  \bibinfo{author}{D.~Lohse}, \bibinfo{author}{X.~Zhang},
\newblock \bibinfo{journal}{Soft Matter} \bibinfo{volume}{13}
  (\bibinfo{year}{2017}) \bibinfo{pages}{2749--2759}.
\bibitem[{Koshkina et~al.(2021)Koshkina, Raju, Kaltbeitzel, Riedinger, Lohse,
  Zhang, and Landfester}]{koshkina2021surface}
\bibinfo{author}{O.~Koshkina}, \bibinfo{author}{L.~T. Raju},
  \bibinfo{author}{A.~Kaltbeitzel}, \bibinfo{author}{A.~Riedinger},
  \bibinfo{author}{D.~Lohse}, \bibinfo{author}{X.~Zhang},
  \bibinfo{author}{K.~Landfester},
\newblock \bibinfo{journal}{ACS Appl. Mater. Interfaces} \bibinfo{volume}{14}
  (\bibinfo{year}{2021}) \bibinfo{pages}{2275--2290}.
\bibitem[{Koos and Willenbacher(2011)}]{koos2011capillary}
\bibinfo{author}{E.~Koos}, \bibinfo{author}{N.~Willenbacher},
\newblock \bibinfo{journal}{Science} \bibinfo{volume}{331}
  (\bibinfo{year}{2011}) \bibinfo{pages}{897--900}.
\bibitem[{Lahnstein et~al.(2008)Lahnstein, Schmehl, R{\"u}sch, Rieger, Seeger,
  and Gessler}]{lahnstein2008pulmonary}
\bibinfo{author}{K.~Lahnstein}, \bibinfo{author}{T.~Schmehl},
  \bibinfo{author}{U.~R{\"u}sch}, \bibinfo{author}{M.~Rieger},
  \bibinfo{author}{W.~Seeger}, \bibinfo{author}{T.~Gessler},
\newblock \bibinfo{journal}{Int. J. Pharm.} \bibinfo{volume}{351}
  (\bibinfo{year}{2008}) \bibinfo{pages}{158--164}.

\end{thebibliography}
\cleardoublepage
\section*{Graphical abstract}
A combination of extraction of the analytes, microdroplet-templated porous supraparticles, and self-lubrication of colloidal ouzo drop evaporation is employed for ultra-sensitive SERS detection. The approach overcomes long-standing limitations of the coffee-stain effect to accumulate the analytes to the hotspots of supraparticles. The ultralow detection of limitation is demonstrated for several model compound solutions and caffeine in saliva samples.
\begin{center}
\includegraphics[height=2in]{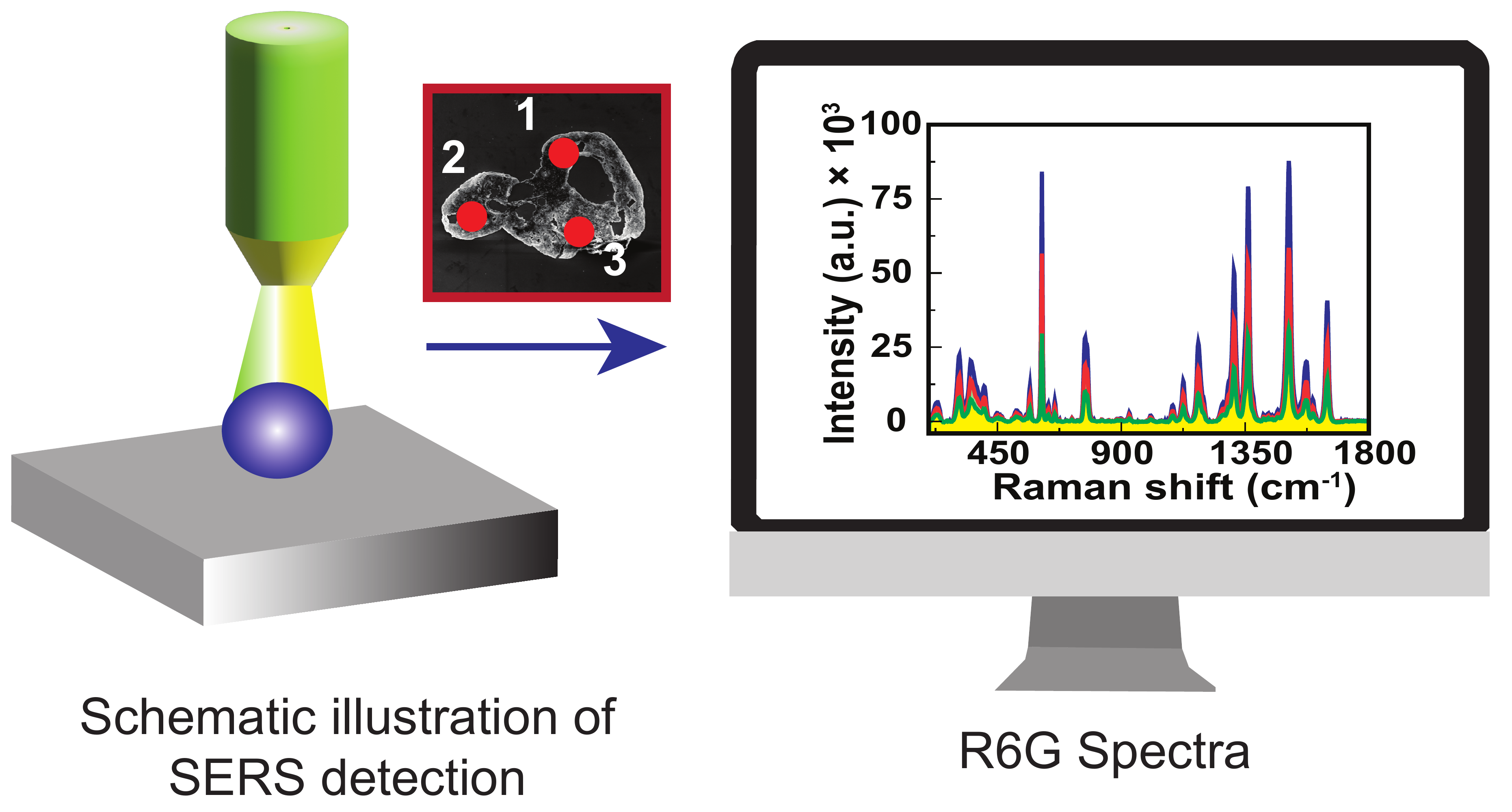}
\end{center}

\cleardoublepage

\end{document}